\documentclass[12pt]{article}
\usepackage[utf8]{inputenc}
\usepackage{amsmath}
\usepackage{amssymb}
\usepackage{a4wide}
\usepackage[usenames,dvipsnames]{color}
\usepackage{verbatim}
\usepackage{mathtools}
\usepackage{mathrsfs}
\usepackage{hyperref}
\usepackage{todonotes}
\usepackage{tikz}
\usepackage{cite}
\usepackage{algorithm}
\usepackage{algorithmicx}
\usepackage{cprotect}
\usepackage{algpseudocode}
\presetkeys{todonotes}{inline}{}
\renewcommand{\d}{\mathrm{d}}

\date{\today}

\begin{document}

\thispagestyle{empty}

\begin{flushright}
  MITP/22-071
% \\ version of \today
\end{flushright}

\vspace{1.5cm}

\begin{center}
  {\Large\bf Amplitudes within causal loop-tree duality\\
  }
  \vspace{1cm}
  {\large Sascha Kromin, Niklas Schwanemann and Stefan Weinzierl \\
  \vspace{1cm}
      {\small \em PRISMA Cluster of Excellence, Institut f{\"u}r Physik, }\\
      {\small \em Johannes Gutenberg-Universit{\"a}t Mainz,}\\
      {\small \em D - 55099 Mainz, Germany}\\
  } 
\end{center}

\vspace{2cm}

% abstract ---------------------------------------
\begin{abstract}\noindent
  {
  We study scalar one-loop amplitudes in massive $\phi^3$-theory within causal loop-tree duality.
  We derive a recurrence relation for the integrand of the amplitude.
  The integrand is by construction free of spurious singularities on H-surfaces.
  Taking renormalisation and contour deformation into account, we obtain the (integrated) amplitude by Monte Carlo integration.
  We have checked up to seven points that our results agree with analytic results.
   }
\end{abstract}
\vspace*{\fill}
\newpage

\section{Introduction}

Loop-tree duality \cite{Catani:2008xa,Bierenbaum:2010cy,CaronHuot:2010zt,Bierenbaum:2012th,Buchta:2014dfa,Hernandez-Pinto:2015ysa,Buchta:2015wna,Sborlini:2016gbr,Driencourt-Mangin:2017gop,Driencourt-Mangin:2019aix,Runkel:2019yrs,Runkel:2019zbm,Baumeister:2019rmh,Aguilera-Verdugo:2019kbz,Driencourt-Mangin:2019yhu,Aguilera-Verdugo:2020set,Plenter:2020lop,Aguilera-Verdugo:2020kzc,Ramirez-Uribe:2020hes,JesusAguilera-Verdugo:2020fsn,TorresBobadilla:2021ivx,Sborlini:2021owe,Bobadilla:2021pvr,deJesusAguilera-Verdugo:2021mvg,Benincasa:2021qcb,Capatti:2019ypt,Capatti:2019edf,Capatti:2020ytd,Capatti:2020xjc,Kermanschah:2021wbk,Capatti:2022tit}
uses Cauchy's residue theorem to reduce each loop momentum integration from a $D$-dimensional integration to a $(D-1)$-dimensional integration, where $D$ denotes the number of space-time dimensions. Typically the energy component of the loop momentum is integrated out with the help of Cauchy's residue theorem. The remaining spatial loop momentum integrations have the same dimensionality as the integrations over the phase space of additional unresolved particles.
Loop-tree duality carries therefore the promise that infrared divergences can be cancelled locally at the integrand level, making the need for any subtraction method obsolete.

However, a naive application of loop-tree duality will generate several terms, where each term
may have new additional spurious singularities. These spurious singularities reside on non-compact hyperbolic hypersurfaces, which are called H-surfaces.
On the other hand, it can be shown that the physical infrared and threshold singularities of an amplitude
reside on compact ellipsoid hypersurfaces. These hypersurfaces are called E-surfaces.

As the singularities on H-surfaces are spurious, they cancel among the terms making up the loop-tree duality representation of an amplitude \cite{Buchta:2014dfa}.
For numerical efficiency and stability it is desirable to make this cancellation manifest. 
The result is called the causal loop-tree duality representation \cite{Aguilera-Verdugo:2019kbz,Aguilera-Verdugo:2020kzc,Capatti:2020ytd}.
Up to now, it is only known how to do this on a graph-by-graph basis, where each Feynman graph generates several
terms in the causal loop-tree duality representation.
An important application of numerical methods like loop-tree duality are of course processes with many particles, where analytical methods run out of steam.
Clearly, a graph-by-graph approach, where in addition each graph generates several terms
is highly inefficient and impractical.

In this paper we investigate the causal loop-tree duality representation directly at the level
of amplitudes. As an example, we study scalar massive $\phi^3$-theory. 
We are interested in the $N$-point one-loop amplitude
\begin{align}
 {\mathscr A}_N^{(1)}\left(p_1,\dots,p_N\right).
\end{align}
After ultraviolet renormalisation, the amplitude is finite. There are no infrared divergences (i.e. there are no infrared $1/\varepsilon$-poles after integration), as all particles are assumed to be massive. However, there will be threshold singularities.
We derive recursion relations for the causal loop-tree duality representation, bypassing the need to talk about Feynman diagrams.

This paper is organised as follows:
In section~\ref{sec:loop_tree_duality} we introduce causal loop-tree duality.
The main result of this paper is derived in section~\ref{sec:recurrence_relations}, where the recurrence relation for the integrand of the amplitude within the causal loop-tree duality representation is worked out.
For numerical results we have to deal with renormalisation and contour deformation, this is done in section~\ref{sec:UV_renormalisation} and section~\ref{sec:contour_deformation}, respectively.
In section~\ref{sec:validation} we validate our approach by comparing for up to seven external particles our results with analytical results.
Our conclusions are given in section~\ref{sec:conclusions}.
In appendix~\ref{sec:app_num_causal_terms} we derive the number of causal terms per diagram.
In appendix~\ref{sec:external_momenta} we give the input data
we used for the numerical checks.

\section{Causal Loop-Tree Duality}
\label{sec:loop_tree_duality}

In this section we review causal loop-tree duality (cLTD).
We start in section \ref{sec:cLTD_representation} with the presentation of the cLTD representation of loop integrals. In section \ref{sec:cLTD_structure_of_the_causal_representation} we investigate the structure of this representation in more detail. A graphical representation of cLTD is introduced in section \ref{sec:graphical_representation}.

\subsection{The causal representation for scalar one-loop diagrams}
\label{sec:cLTD_representation}

We consider a scalar one-loop integral $I$, given by
\begin{align}
\label{eq:one_loop_scalar_integral}
    I = \int \frac{\mathrm{d}^D k}{(2\pi)^D} \frac{1}{\prod_{j=1}^N \left[(k+q_j)^2 - m_j^2 + i\delta\right]},
\end{align}
 with loop momentum $k$ and $N$ external particles. The momenta of the external particles are denoted by $p_1, \dots, p_N$. 
\begin{figure}
\begin{center}
\includegraphics[scale=1.0]{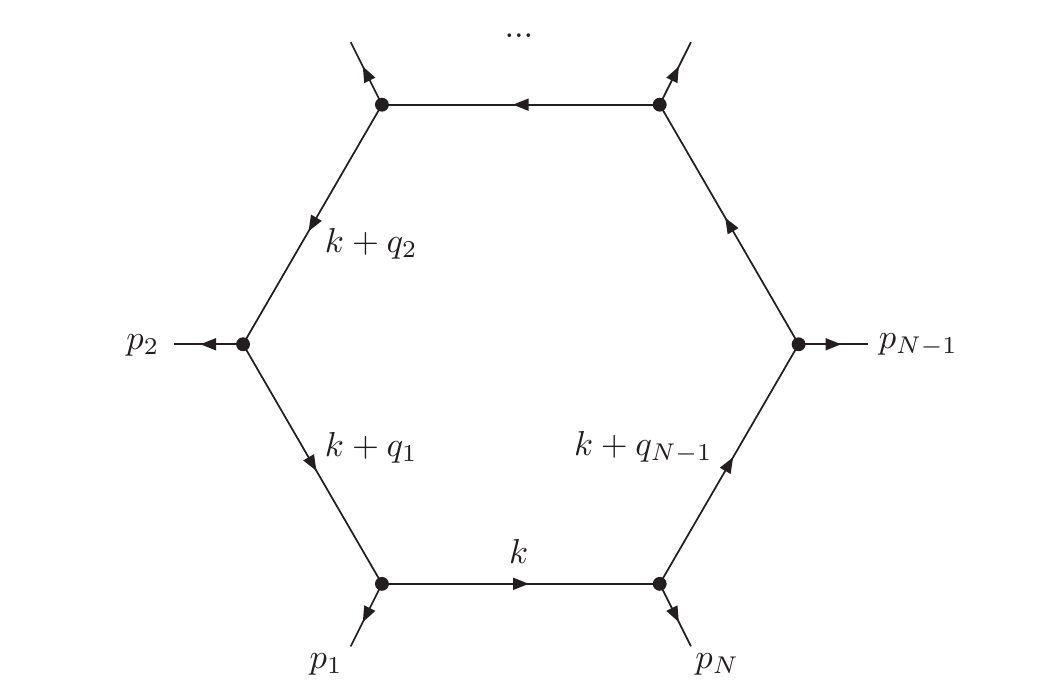}
\end{center}
\caption{\small
The labelling for a generic one-loop integral. The arrows denote the momentum flow.
}
\label{fig:kinematics}
\end{figure}
The kinematics is shown in fig.~\ref{fig:kinematics}.
The external particles are taken as outgoing and satisfy momentum conservation
\begin{align*}
	\sum_{i=1}^N p_i = 0.
\end{align*}
The mass of propagator $j$ is given by $m_j$ and the quantities $q_j$ are linear combinations of external momenta,
\begin{align}
    \label{eq:simple_lin_comb_ext_moms}
	q_j \coloneqq \sum_{i=1}^j p_i.
\end{align}
We also define a quantity for the difference of two propagator momenta,
\begin{align*}
	q_{ij} \coloneqq q_i - q_j.
\end{align*}
By applying the residue theorem to eq. \eqref{eq:one_loop_scalar_integral}, we obtain the loop-tree duality representation \cite{Catani:2008xa} of the integral $I$,
\begin{align*}
    I = -i\int \frac{d^{D-1} k}{(2\pi)^{D-1}} \sum_{i=1}^N \frac{1}{2E_i} \frac{1}{\prod_{\substack{j=1,\\j\neq i}}^N\left[(E_i - q_{ij}^0)^2 - E_j^2\right]}, \hspace{0.3cm} \text{where} \hspace{0.2cm} E_i = \sqrt{\left(\Vec{k} + \Vec{q}_i\right)^2 + m_i^2 - i\delta}.
\end{align*}
The propagators that are not set on-shell by taking the residue are called \textit{dual propagators} because they exhibit a different $i\delta$ prescription than the original propagators.
As already extensively discussed in the literature, see \textit{e.g.} \cite{Buchta:2014dfa, Buchta:2015jea, Hernandez-Pinto:2015ysa, Buchta:2015wna, Sborlini:2016gbr, Capatti:2019ypt}, the dual propagators can become singular on two different kinds of surfaces, which are called H-surfaces and E-surfaces. The singularities on H-surfaces are spurious and cancel analytically, although these so called \textit{dual cancellations} can lead to numerical instabilities. For this reason a different representation, the causal loop-tree duality representation \cite{Capatti:2020ytd,Aguilera-Verdugo:2020kzc}, is an interesting alternative.
In order to define causal loop-tree duality let us consider the expressions
\begin{align}
	\pm E_i \pm E_j - q_{ij}^0.
\end{align}
Alternating signs in front of the energies define H-surfaces, equal signs in front of the energies define E-surfaces.
Geometrically, H-surfaces correspond to the intersection of two forward hyperboloids
or two backward hyperboloids, while E-surfaces correspond to the intersection of a 
forward hyperboloid with a backward hyperboloid.
We set
\begin{align}
    \label{eq:E-surface}
	x_{ij} &\coloneqq E_i + E_j - q_{ij}^0.
\end{align}
Each $x_{ij}$ defines an E-surface.
In the causal loop-tree duality representation only the $x_{ij}$'s (and the energy factors $E_i$) appear as factors in the denominator.
\\
In a similar approach to \cite{Capatti:2020ytd}, we independently derived a locally equivalent expression for the causal representation for scalar one-loop diagrams,
\begin{align}
	\label{eq:final_result_scalar_avg}
	\begin{split}
	    \int \frac{\d^D k}{(2\pi)^D} &\frac{1}{\prod_{j = 1}^N[(k+q_j)^2 - m_j^2 + i\delta]}\\
    	&= i \int \frac{\d^{D-1} k}{(2\pi)^{D-1}} \frac{(-1)^{N}}{\prod_{i=1}^N \left(2E_i\right)} \sum_{\substack{N_L=1,\\N_R=N-N_L}}^{N-1} \sum_{A_L \sqcup A_R} \mathcal{F} \left(A_L, A_R, \{x_{ij}\}\right),
	\end{split}
\end{align}
where we introduced a function $\mathcal{F}$ given by
\begin{align}	
    \label{eq:pprod}
    \begin{split}
        \phantom{=} \mathcal{F}(A_L, A_R, \{x_{ij}\}) \coloneqq \sum_{m_1=1}^{N_R}\frac{1}{\prod_{n_1=m_1}^{N_R} x_{(\ell_1) (r_{n_1})}} \cdots \sum_{m_{N_L-2}=1}^{m_{N_{L-3}}} \frac{1}{\prod_{n_{N_L-2}=m_{N_L-2}}^{m_{N_L-3}} x_{(\ell_{N_L-2}) (r_{n_{N_L-2}})}} \times\\
    	\times \sum_{m_{N_L-1}=1}^{m_{N_L-2}} \frac{1}{\left[\prod_{n_{N_L-1}=m_{N_L-1}}^{m_{N_L-2}} x_{(\ell_{N_L-1}) (r_{n_{N_L-1}})}\right]\left[\prod_{n_{N_L}=1}^{m_{N_L-1}} x_{(\ell_{N_L}) (r_{n_{N_L}})}\right]}.
    \end{split}
 \end{align}
The second sum in eq. \eqref{eq:final_result_scalar_avg} runs over all partitions of the ordered set $A \coloneqq \{1,\dots,N\}$ into two disjoint ordered subsets $A_L$ and $A_R$ with cardinalities $N_L$ and $N_R$, respectively. The cardinalities are determined by the first sum in the expression. Both ordered subsets inherit their order from $A$ and are given by $A_L = \{\ell_1, \dots, \ell_{N_L}\}$ and $A_R = \{r_1, \dots, r_{N_R}\}$. The representation given in eq. \eqref{eq:final_result_scalar_avg} is not unique but instead depends on the ordering of the elements of $A$. A graphical representation of the function $\mathcal{F}$ is given in section \ref{sec:graphical_representation}.\\
For the indices of the $x_{ij}$'s we have
\begin{align}
    \label{eq:E-surface_indices}
	 i \in A_L, &\;\;\; j \in A_R.
\end{align}
To get more comfortable with the function $\mathcal{F}$, we list examples for $A_L = \{\ell_1\}$, $A_L = \{\ell_1, \ell_2\}$ and $A_L = \{\ell_1, \ell_2, \ell_3\}$,
\begin{align}
	\label{eq:pprod_first_term}
	\begin{split}
	    \mathcal{F}(\{\ell_1\}, A_R, \{x_{ij}\}) &= \frac{1}{\prod_{n_1=1}^{N_R} x_{(\ell_1) (r_{n_1})}},\\
    	\mathcal{F}(\{\ell_1, \ell_2\}, A_R, \{x_{ij}\}) &= \sum_{m_1=1}^{N_R} \frac{1}{\left[\prod_{n_1=m_1}^{N_R} x_{(\ell_1) (r_{n_1})}\right]\left[\prod_{n_2=1}^{m_1} x_{(\ell_2) (r_{n_2})}\right]},\\
    	 \mathcal{F}(\{\ell_1, \ell_2, \ell_3\}, A_R, \{x_{ij}\}) &= \sum_{m_1=1}^{N_R} \frac{1}{\prod_{n_1=m_1}^{N_R} x_{(\ell_1) (r_{n_1})}} \sum_{m_2=1}^{m_1} \frac{1}{\left[\prod_{n_2=m_2}^{m_1} x_{(\ell_2) (r_{n_2})}\right]\left[\prod_{n_3=1}^{m_2} x_{(\ell_3) (r_{n_3})}\right]}.
	\end{split}
\end{align}
The parentheses that distinguish the indices on the E-surfaces introduced in this section only serve the readability and will be omitted in what follows.

\subsection{Structure of the causal representation}
\label{sec:cLTD_structure_of_the_causal_representation}
In this subsection we want to investigate the cLTD representation in more detail. Therefore we present some explicit examples in the following.
Let us start by presenting the causal representation of the $2$-point function,
\begin{align*}
    \int \frac{\d^D k}{(2\pi)^D} \frac{1}{\prod_{j=1}^2[(k+q_j)^2 - m_j^2 + i\delta]} = i \int \frac{\d^{D-1} k}{(2\pi)^{D-1}} \frac{1}{4E_1E_2} \left(\frac{1}{x_{12}} + \frac{1}{x_{21}}\right).
\end{align*}
At this point we like to remind that the terms $x_{12}$ and $x_{21}$ are not the same, since the defining equations for the E-surfaces are not symmetric in the indices, \textit{c.f.} eq. \eqref{eq:E-surface}.
For the $3$-point function we have,
\begin{align*}
\lefteqn{
	\int \frac{\d^D k}{(2\pi)^D} \frac{1}{\prod_{j=1}^3[(k+q_j)^2 - m_j^2 + i\delta]} }\\
	& = -i \int \frac{\d^{D-1} k}{(2\pi)^{D-1}} \frac{1}{8E_1E_2E_3} \left[\frac{1}{x_{13} x_{12}} + \frac{1}{x_{23} x_{21}} + \frac{1}{x_{32} x_{31}} + \frac{1}{x_{13} x_{23}} + \frac{1}{x_{12} x_{32}} + \frac{1}{x_{21} x_{31}}\right],
\end{align*}
and the causal representation of the $4$-point function is given by
\begin{align}
    \label{eq:cLTD_box_std}
    \lefteqn{
	\int \frac{\d^D k}{(2\pi)^D} \frac{1}{\prod_{j=1}^4[(k+q_j)^2 - m_j^2 + i\delta]} }\\
&
\begin{split}
= i \int \frac{\d^{D-1} k}{(2\pi)^{D-1}} \frac{1}{16E_1E_2E_3E_4}\left[\frac{1}{x_{14} x_{13} x_{12}} + \frac{1}{x_{24} x_{23} x_{21}} + \frac{1}{x_{34} x_{32} x_{31}} + \frac{1}{x_{43} x_{42} x_{41}} \right.\\
	\left. + \frac{1}{x_{14} x_{13} x_{23}} + \frac{1}{x_{14} x_{24} x_{23}} + \frac{1}{x_{14} x_{12} x_{32}} + \frac{1}{x_{14} x_{34} x_{32}}\right.\\
	\left. + \frac{1}{x_{13} x_{12} x_{42}} + \frac{1}{x_{13} x_{43} x_{42}} + \frac{1}{x_{24} x_{21} x_{31}} + \frac{1}{x_{24} x_{34} x_{31}} \right.\\
	\left. + \frac{1}{x_{23} x_{21} x_{41}} + \frac{1}{x_{23} x_{43} x_{41}} + \frac{1}{x_{32} x_{31} x_{41}} + \frac{1}{x_{32} x_{42} x_{41}} \right.\\
	\left. + \frac{1}{x_{14} x_{24} x_{34}} + \frac{1}{x_{13} x_{23} x_{43}} + \frac{1}{x_{12} x_{32} x_{42}} + \frac{1}{x_{21} x_{31} x_{41}} \right].
	\end{split}
	\nonumber
\end{align}
By investigating the examples given above one can see that the causal representation consists of a product of on-shell energies that multiplies a sum of terms whose denominator is a monomial in the variables $x_{ij}$.  
The number of terms of an $N$-point function for $N \geq 2$ is given by the central binomial coefficient,
\begin{align}
    \label{eq:number_of_causal_terms_diagram}
    %\sum_{j=1}^{N-1}\binom{N}{j} \binom{N-2}{j-1}.
    N_\text{terms} = \binom{2(N-1)}{N-1}.
\end{align}
The derivation of the above expression is given in appendix \ref{sec:app_num_causal_terms}. However, all terms in an $N$-point function can be constructed from a small number of different E-surfaces, given by
\begin{align}
    \label{eq:number_of_E_surfaces_diagram}
    N_\text{E-surfaces} = 2\cdot \binom{N}{2},
\end{align}
making recurrence relations a favourable approach for numerical evaluations. This can easily be seen by noticing that an E-surface is described by two indices. Hence, the number of E-surfaces is given by the number of possibilities to draw two indices out of $N$. The factor of two originates from the fact that the $x_{ij}$ are not symmetric in $i$ and $j$ and therefore the order in which the indices are drawn matters. Table \ref{tab:num_terms_vs_num_Esurfaces_function} shows the number of terms of the cLTD representation and the number of E-surfaces for $N$-point functions with $2 \leq N \leq 8$.\\
\begin{table}[ht]
    \centering
    \begin{tabular}{c|c|c|c|c|c|c|c}
        $N$ & $2$ & $3$ & $4$ & $5$ & $6$ & $7$ & $8$\\\hline\hline
        Number of terms & $2$ & $6$ & $20$ & $70$ & $252$ & $924$ & $3432$\\\hline
        Number of E-surfaces & $2$ & $6$ & $12$ & $20$ & $30$ & $42$ & $56$
    \end{tabular}
    \caption{\small Comparison of the number of terms and the number of E-surfaces of the cLTD representation of an $N$-point function.}
    \label{tab:num_terms_vs_num_Esurfaces_function}
\end{table}
In order to construct recurrence relations based on the cLTD representation we first need to understand how to construct valid terms. Therefore, we need to find criteria that have to be satisfied for a term to be valid. Finding these criteria is what we will do in the following.\\
If one carefully investigates the examples given at the beginning of this subsection, one finds that each summand contains all possible indices, and an index can never be on the left- and on the right-hand side in the same summand. This is manifested in eq. \eqref{eq:final_result_scalar_avg}, where the sum $\sum_{A_L \sqcup A_R}$ runs over all disjoint subsets whose union equals the set of all indices.\\
We can obtain more information about the structure of causal terms by considering the right-hand side of equation \eqref{eq:pprod}. Remember that the elements of $A_L$ are denoted $\ell_1, \dots, \ell_{N_L}$ and the elements of $A_R$ by $r_1, \dots, r_{N_R}$. In each product in the denominator the index $i$ of $\ell_i \in A_L$ is raised by one, \textit{i.e}, in the first product we have $\ell_1$, in the second one $\ell_2$ and in the $j^{\text{th}}$ product we have $\ell_j$ as left index of the E-surface. The products run over the right index $r \in A_R$ so that the $i^{\text{th}}$ product runs from $m_i$ to $m_{(i-1)}$, \textit{i.e.}, $m_i \leq m_{(i-1)}$. The values of $m_i$ and $m_{(i-1)}$ are determined by the sums in eq. \eqref{eq:pprod} and are not important for the current discussion. Consequently, the right index in the $i^{\text{th}}$ product runs from $r_{m_i}$ to $r_{m_{(i-1)}}$, so that we obtain the E-surfaces $x_{\ell_ir_{m_i}}, \dots, x_{\ell_ir_{m_{(i-1)}}}$ as a result of this product. If we now take a look at the previous product, we see that the right index runs from $r_{m_{(i-1)}}$ to $r_{m_{(i-2)}}$ yielding the E-surfaces $x_{\ell_{(i-1)}r_{m_{(i-1)}}}, \dots, x_{\ell_{(i-1)}r_{m_{(i-2)}}}$. Thus, the largest right index of an E-surface appearing in the $i^{\text{th}}$ product, or equivalently, accompanying the index $\ell_i$, is the same index as the smallest index accompanying the index $\ell_{(i-1)}$, which is in both cases the index $r_{m_{(i-1)}}$.\\
Since this is still rather abstract, let us illustrate this in an example. Consider a term of the $6$-point function,
\begin{align}
    \label{eq:example_six_point_term}
    \frac{1}{x_{16}x_{15}x_{35}x_{32}x_{42}}.
\end{align}
For this term we have $A_L = \{1,3,4\}$ and $A_R = \{2,5,6\}$ and we can see that the first two criteria are satisfied. Now let us consider $\ell_2 = 3$. The largest index $r \in A_R$ accompanying the index $\ell_2$ is $r_2 = 5$. The index $\ell_1$ that denotes the next smaller index in $A_L$ is $1$. We see that the smallest index $r \in A_R$ accompanying $1$ is again $r_2 = 5$, which is what we wanted. This must also be true for $\ell_3 = 4$. Here we have $r_1 = 2$, which is also the smallest index appearing in combination with $\ell_2 = 3$. Thus, the term in eq. \eqref{eq:example_six_point_term} is a valid term.\\
To conclude this section, let us summarise the conditions for valid terms:
\begin{itemize}
    \item[1)] All indices of the defining set $A$ must occur in the term, \textit{i.e.}, $A_L \cup A_R = A$.
    \item[2)] Each index can only appear on one side, \textit{i.e.}, $A_L \cap A_R = \emptyset$.
    \item[3)] The largest index accompanying an index $\ell_i \in A_L$ must be the smallest index accompanying the index $\ell_{(i-1)} \in A_L$.
\end{itemize}

\subsection{Graphical representation}
\label{sec:graphical_representation}
\begin{figure}[ht!]
    \centering
    \includegraphics[width=1\textwidth]{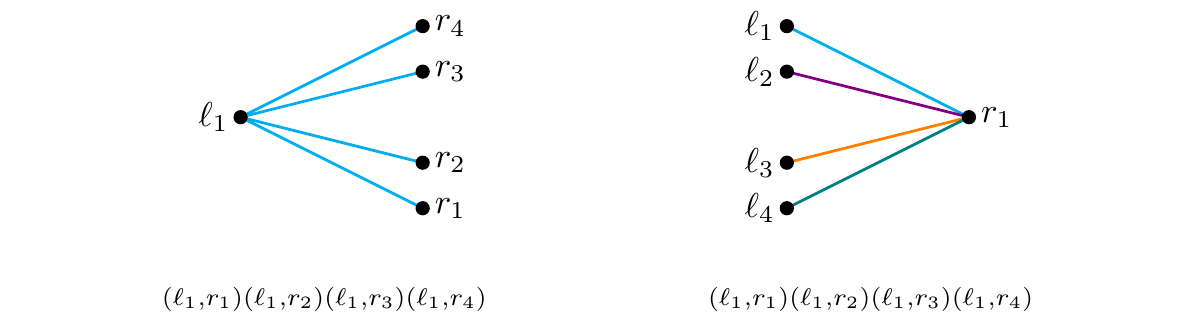}
    \includegraphics[width=1\textwidth]{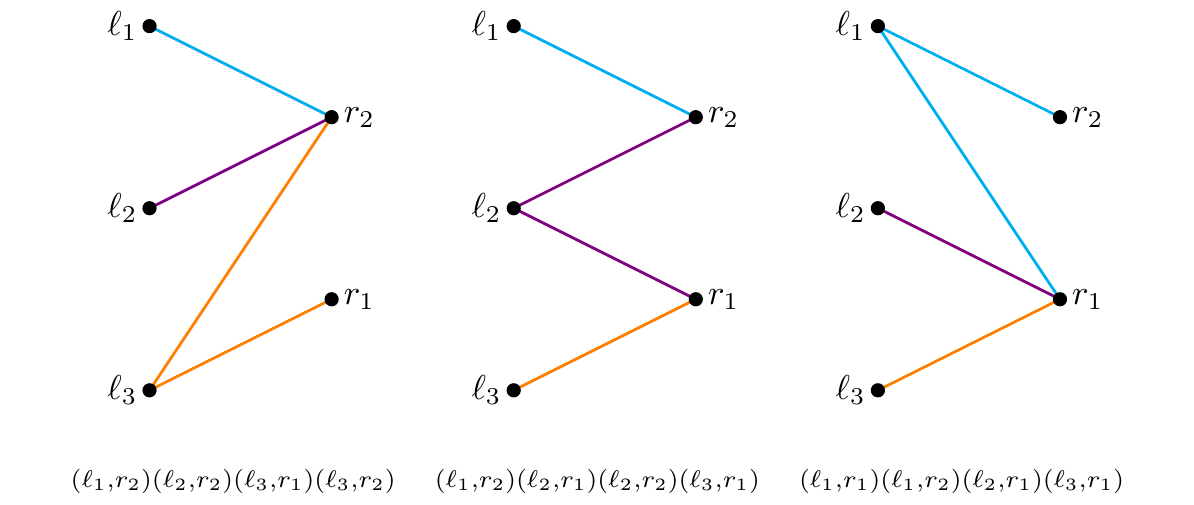}
    \includegraphics[width=1\textwidth]{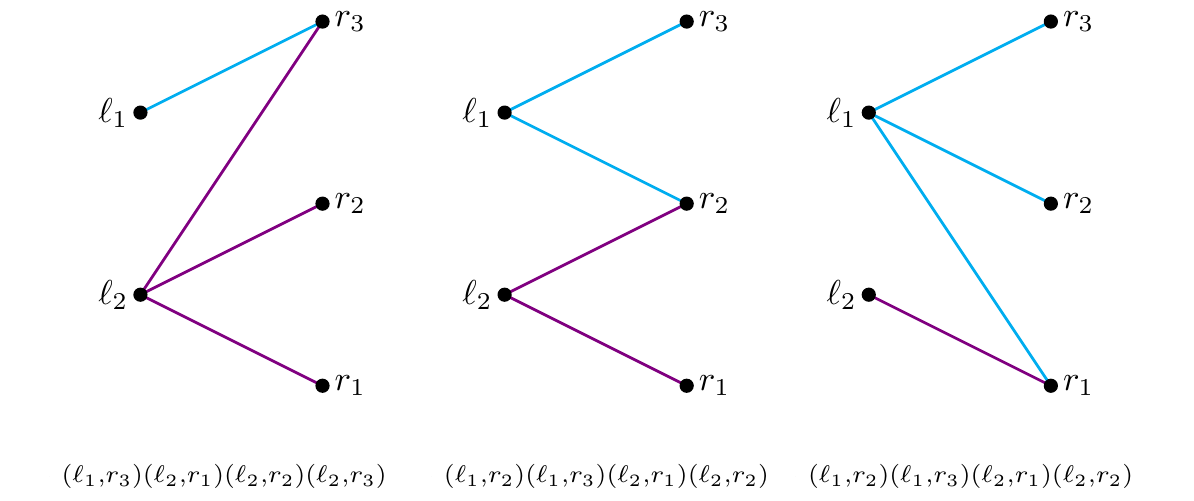}
    \caption{\small The eight different topologies of causal diagrams of the five-point function.}
    \label{fig:F_representation}
\end{figure}

The purpose of this section is to give a graphical representation of the function $\mathcal{F}$ defined in eq. \eqref{eq:pprod}. Before we start, let us introduce a more compact notation for causal terms,
\begin{align*}
    \frac{1}{x_{\ell_1 r_j} \dots x_{\ell_{i}r_{k}}} \equiv (\ell_1,r_j) \cdots (\ell_{i}, r_{k}),
\end{align*}
which we will use in the following.\\
The important quantities that we like to present in this section will be called causal diagrams. A causal diagram consists of left and right vertices which correspond to elements of the sets $A_L$ and $A_R$, respectively. Connecting two vertices defines an E-surface $(\ell,r)$, where $\ell \in A_L$ corresponds to a left vertex and $r \in A_R$ to a right vertex. The most basic diagram consisting of two vertices, representing one E-surface, is shown in figure \ref{fig:causal_diagram_E-surface}.
\begin{figure}[ht]
\centering
\begin{tikzpicture}
\begin{scope}[scale=1]

%Loop
\draw[gray, thick] (0,0) -- (1,0);

\coordinate(1) at (-0.2,0);
\coordinate(2) at (1.2,0);

\node at (1) {$\ell$};
\node at (2) {$r$};

\end{scope}
\end{tikzpicture}
\caption{\small Causal diagram representing a single E-surface $(\ell,r)$}
\label{fig:causal_diagram_E-surface}
\end{figure}
There are $\binom{N-2}{N_L-1}$ different topologies of causal diagrams each containing $\binom{N}{N_L}$ different diagrams corresponding to the different permutations of indices. 
A fixed topology together with a set $A_L$ results in a unique causal diagram and hence a unique term.\\
We will now give instructions on how to draw all causal diagrams for given sets $A_L$ and $A_R$:
\begin{itemize}
    \item[1)] Draw $N_L$ vertices on the left-hand side and $N_R$ vertices on the right-hand side of the diagram.
    \item[2)] Label the vertices on the left-hand side with the indices of $A_L$ by respecting their order and do the same with the vertices on the right-hand side, but with the reversed order of $A_R$.
    \item[3)] Connect the left and right vertices by straight lines, such that:
    \begin{itemize}
        \item[a)] Each vertex has at least valency one
        \item[b)] A line always connects exactly one left vertex with one right vertex.
        \item[c)] There are no intersections among the lines.
    \end{itemize}
\end{itemize}
Point 3a) follows from the fact that all indices of the ordered set $A$ must occur in a causal term, \textit{cf.} rule $1)$ in section \ref{sec:cLTD_structure_of_the_causal_representation}. The fact that each E-surface carries one left index $\ell \in A_L$ and one right index $r \in A_R$ is reflected in point 3b). The last point 3c) is a direct consequence of rule $3)$ in section \ref{sec:cLTD_structure_of_the_causal_representation}.\\
As an explicit example consider $A_L = \{\ell_1, \ell_2\}$ of the five point function. Following the rules we have $\binom{3}{1} = 3$ different topologies of causal diagrams shown at the bottom of figure \ref{fig:F_representation}.
Each diagram contributes with $\binom{5}{2} = 10$ terms, corresponding to all possibilities of selecting two elements $\ell_1$ and $\ell_2$ from the five elements of $A$.
The three different types of diagrams represent the following terms,
\begin{align}
    (\ell_1,r_3)(\ell_2,r_1)(\ell_2,r_2)(\ell_2,r_3), \;(\ell_1,r_2)(\ell_1,r_3)(\ell_2,r_1)(\ell_2,r_2), \;(\ell_1,r_1)(\ell_1,r_2)(\ell_1,r_3)(\ell_2,r_1).
\end{align}
For completeness, all possible topologies of the 5-point functions are shown in figure \ref{fig:F_representation}.

\section{Recurrence relations}
\label{sec:recurrence_relations}

In this section we present our main results, a recursion formula for scalar one-loop amplitudes within the cLTD formalism. The goal of this section is to introduce an alternative to the graph-by-graph approach for constructing a scalar one-loop amplitude within the cLTD formalism. Throughout this section we will use the convention $2 \leq n \leq N$, where $n$ refers to an $n$-point function while $N$ refers to the number of external particles. 
This describes the situation, where $(N-n)$ out of the $N$ possible loop propagators are pinched.
We will use the notation for causal terms that was introduced in section \ref{sec:graphical_representation}.

\subsection{Structure of the amplitude}
\label{sec:structre_of_the_amplitude}

Our standard examples will be one-loop $N$-point amplitudes in scalar $\phi^3$-theory.
These amplitudes are not cyclic ordered.
However, there is a dihedral symmetry at the level of the Feynman diagrams, generated by the cyclic permutations and the reflection.
We therefore obtain $(N-1)!/2$ possible cyclic orders of the external legs. 
We split the amplitude into contributions with a definite cyclic order.
The reason is the following: In section~\ref{sec:contour_deformation} we define for each cyclic order an individual contour deformation.
We integrate each cyclic order with this individual integration contour.
This is very natural in QCD and Yang-Mills theory, where the cyclic order is induced by colour ordering. For a scalar $\phi^3$-theory it does no harm to introduce the cyclic order by hand.
Let us point out that for each cyclic order
we have exactly one $N$-point function.
\\
Since the procedure of the algorithm is the same for each cyclic order of the external legs, we will focus on one cyclic order in what follows.\\
The ordered set $A$ introduced in section \ref{sec:cLTD_representation} will refer from now on to the $N$-point function in the cyclic order under consideration. We will denote the elements $A$ by $a_i \coloneqq \sigma(i) \bmod N$, where $\sigma(i)$ is the corresponding cyclic permutation of the index $i$. It is advantageous starting the counting with $0$, i.e. $a_j \in \{0,\dots,N-1\}$. The elements are ordered by their indices such that the ordered set reads $A = \{a_1,\dots,a_N\}$.\\
Let us now investigate the structure of the one-loop $N$-point amplitude. We will use the convention that the smallest index is always $a_1 = 0$ since one can always shift the momenta accordingly and use momentum conservation to eliminate an index. This is also true for all $n$-point functions (with $n\le N$). Combining this with the fact that the causal terms of an $n$-point function carry exactly $n$ indices, an $n$-point function is specified by the remaining $(n-1)$ indices. This leaves $\binom{N-1}{n-1}$ different combinations of indices for an $n$-point function, and thus, $\binom{N-1}{n-1}$ $n$-point functions per cyclic order.\\
This ansatz for the construction of the amplitude introduces double counting between different cyclic orders that has to be accounted for by symmetry factors, which will be discussed in section \ref{sec:recurrence_relations_sym_facs}.
In the following we consider the standard order of the external legs.

\subsection{Recursive construction of the integrand}
Due to the exponential growth of terms of the cLTD integrands with the number of external legs, we aim for a method where we do not need to generate each term individually. The algorithm essentially consists of two steps. 
In the first step we generate a set of simple building blocks, which we call base terms.
The base terms may have fewer indices.
In the second step we dress recursively the terms obtained so far with additional indices.
The recursive step generates a causal representation of a higher-point function from a lower-point function.

For each term of a lower-point function we introduce an ordered set $A^\flat$ containing all of its indices. The ordered set $A^\flat$ is the analogue of $A$ for base terms. The sets $A_L^\flat$ and $A_R^\flat$ are defined analogous to $A_L$ and $A_R$, so that $A_L^\flat \sqcup A_R^\flat = A^\flat$. Elements in these sets are denoted by $\ell_i^\flat \in A_L^\flat$ and $r_i^\flat \in A_R^\flat$.\\
In section \ref{sec:uniqueness_of_terms} we show that for a given cyclic order we generate each causal term only once, so that we do not need to introduce symmetry factors within a given cyclic order.\\
Throughout this section we will only consider the case where $\ell_1^\flat = 0$, \textit{i.e.}, $0 \in A_L^\flat$. 
The terms with $0 = r_1^\flat \in A_R^\flat$ can be obtained analogously by exchanging the indices $\ell$ and $r$ in the following.\\
Furthermore we will use the sets $A_L$ and $A_R$ rather symbolically to refer to an index that will be added as a left or a right index to the base term, respectively.\\

\subsubsection{Generation of causal terms}
\label{sec:generation_of_causal_terms}
The heart of the algorithm is the recursive construction of causal terms starting from the base terms.
In the first part of this section we describe the structure of the base terms, which we will prove in section \ref{sec:recursion}. We then construct new terms by adding new vertices to causal diagrams by respecting the rules established at the end of section \ref{sec:cLTD_structure_of_the_causal_representation}.

\subsubsection*{Base terms}
\begin{figure}[ht!]
    \centering
    \includegraphics[width=0.9\textwidth]{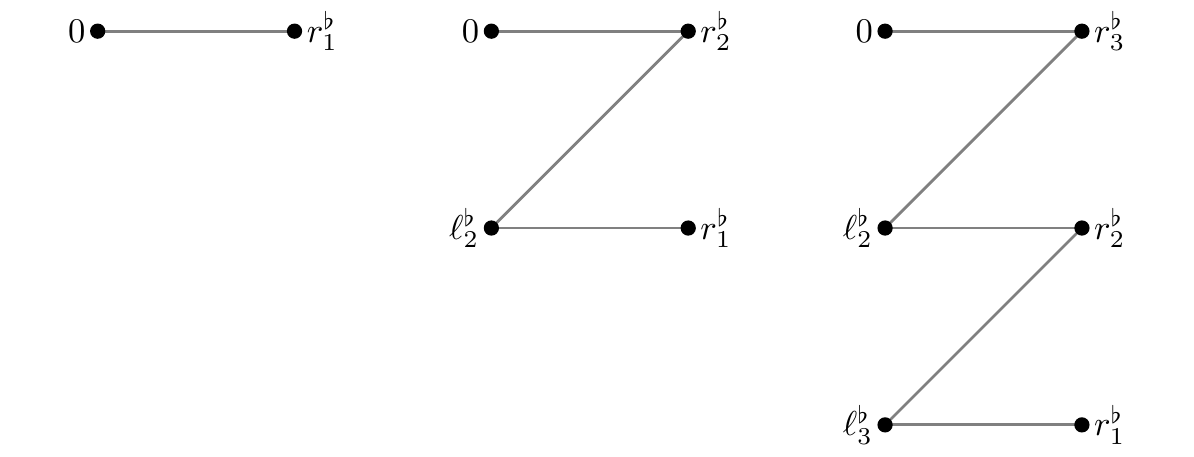}
    \caption{\small Base term causal diagram topology with fixed $\ell_1^{\flat}=0$ and $m=1,2,3$ from left to right.}
    \label{fig:base_terms_causal_diagrams}
\end{figure}

The base terms have the general form
\begin{align}
    \label{eq:base_term_recursion}
    (0,r_m^\flat)\left[\prod_{i=2}^m (\ell_i^\flat, r_{m-i+1}^\flat) (\ell_i^\flat, r_{m-i+2}^\flat)\right],
\end{align}
which we prove in section~\ref{sec:recursion}. 
Here, we made use of the identification $\ell_1^\flat = 0$. The index $m$ indicates the cardinality of $A_R^{\flat}$ and $A_L^{\flat}$, so that $A^\flat$ contains $2m$ elements. In other words, the base terms can only be those terms that originate from Feynman diagrams with an even number of external legs, \textit{e.g.} $m=1$ corresponds to bubbles, $m=2$ to boxes \textit{etc.}\\
Base terms correspond to a specific "zigzag"-topology of causal diagrams, shown in figure \ref{fig:base_terms_causal_diagrams} for the first examples $m = 1,2,3$.

Let us now discuss the first recursive step from a base term with $n$ points and indices $A^{\flat}$ to $(n+1)$ points.
In order to generate a valid $(n+1)$-point term with the expression in \eqref{eq:base_term_recursion}, we need an E-surface with an index $a_k \in A$ and $a_k \notin A^\flat$ as a consequence of rule $1)$. Condition $2)$ allows us to construct terms with either $a_k \in A_L$ or $a_k \in A_R$. These two cases are slightly different due to the fact that we set $\ell_1^\flat = 0$.
\subsubsection*{Adding a left vertex}
Let us start with the case $a_k \in A_L$.
\begin{figure}[ht!]
    \centering
    \includegraphics[width=0.9\textwidth]{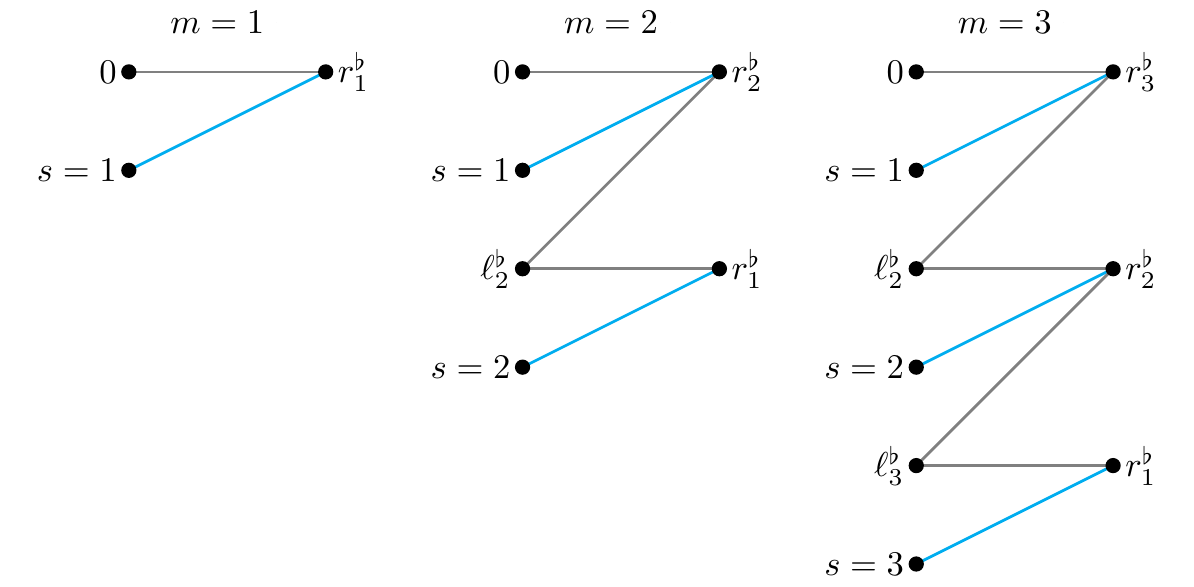}
    \caption{\small Causal diagram for adding a left vertex to base terms (black lines) for $m=1,2,3$ from left to right. The blue lines illustrate factors that can be multiplied to the base term to yield a valid term, resulting in a new topology of the causal diagram. Not all possibilities of adding a blue line might be allowed. This depends on the set $A_L^\flat$.}
    \label{fig:adding_left_vertex}
\end{figure}
In order to fulfil condition $3)$, we have to know how many indices $\ell_i \in A_L^\flat$ exist, with $i = 1,\dots, N_L^\flat$, that are smaller than $a_k$.
Let us denote this number by $s$ and let $s > 0$, so that $\ell_s^\flat < a_k < \ell_{s+1}^\flat$, or just $\ell_s^\flat < a_k$ if $s = N_L^\flat$.
This is illustrated in figure \ref{fig:adding_left_vertex} for $m \in \{1,2,3\}$. The black lines connect the vertices that belong to the base term, while the blue lines lead to new vertices.
The labels on the vertices of valency one of these blue lines tell us how many left vertices are above the new one, \textit{i.e.}, how many $\ell_i \in A_L^\flat$ exist that are smaller than $a_k$.\\
Given condition 3) or figure \ref{fig:adding_left_vertex}, the index $r^\flat \in A_R^\flat$ accompanying $a_k$ can now be determined.
The smallest index accompanying $\ell_s^\flat$ is $r_{m-s+1}^\flat$, as can be seen from expression \eqref{eq:base_term_recursion}.
As a consistency check one can also take a look at the largest index accompanying $\ell_{s+1}^\flat$, if it exists, which is also given by $r_{m-s+1}^\flat$. 
Thus, we can multiply the term 
\begin{align}
    \label{eq:rec_ak_left}
    (a_k,r_{m-s+1}^\flat)
\end{align}
to expression \eqref{eq:base_term_recursion}. 
For $s = 0$ the term in equation \eqref{eq:rec_ak_left} does not exist since then we would have $m-s+1 = m+1 > m = N_L^{\flat}$. Luckily, this does not lead to a problem because $s=0$ corresponds to the case where the new index $a_k$ is smaller than all elements of $A_L^\flat$. However, we set $\ell_1^\flat = 0$, so this case cannot occur. 

\subsubsection*{Adding a right vertex}
\begin{figure}[ht!]
    \centering
    \includegraphics[width=0.9\textwidth]{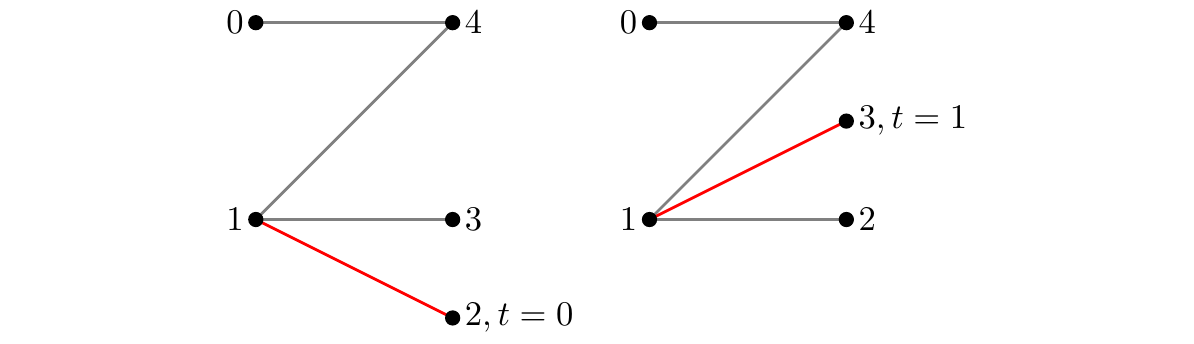}
    \caption{\small Causal diagram for adding a right vertex to base terms with $t=0$ and $t=1$, which produce the same diagram.}
    \label{fig:t=0_illustration}
\end{figure}

Now consider the case $a_k \in A_R$.
\begin{figure}[ht!]
    \centering
    \includegraphics[width=0.9\textwidth]{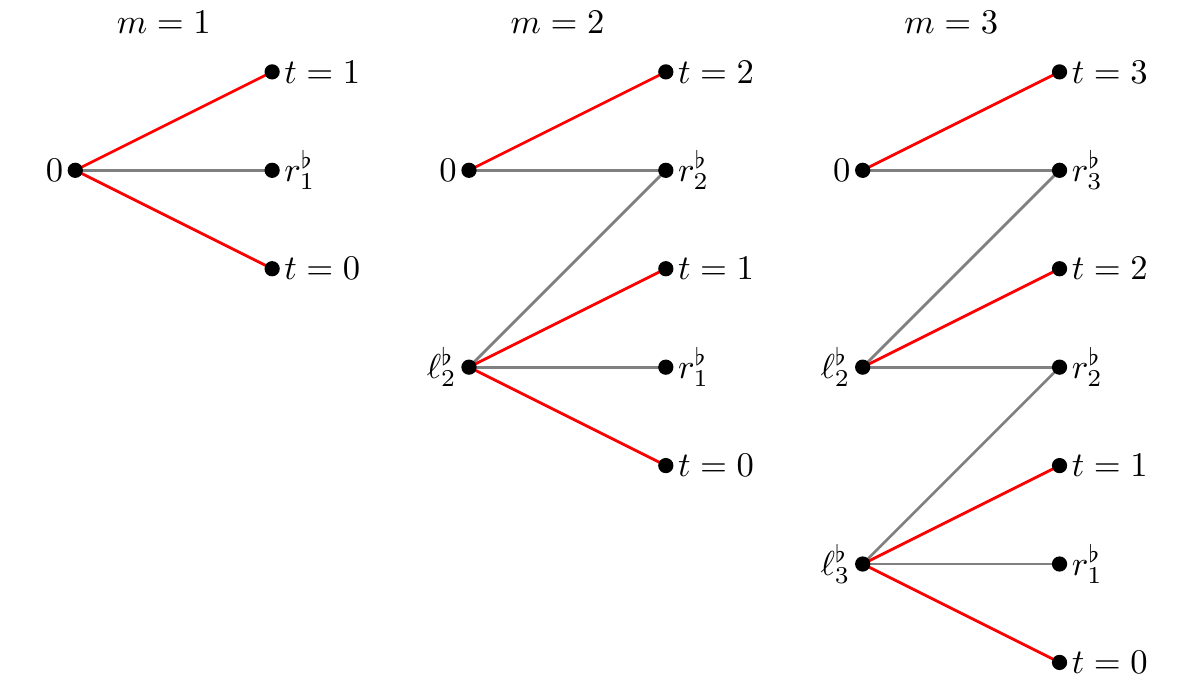}
    \caption{\small Same as figure \ref{fig:adding_left_vertex}, but for the introduction of a new right vertex. The lines connecting the new vertex with the respective base term diagram are shown in red.}
    \label{fig:adding_right_vertex}
\end{figure}
Let the number of indices, $r_i^\flat \in A_R^\flat$, that are smaller than $a_k$ be given by $t$, so that $r_t^\flat < a_k < r_{t+1}^\flat$, or just $r_t^\flat < a_k$ if $t = N_R^\flat$.
Let us start again with the case that $t > 0$. In order to construct a valid term with $a_k \in A_R$ we note that each right index $r_i^\flat \in A_R^\flat$ in expression \eqref{eq:base_term_recursion} appears twice, once as smallest and once as largest index of two subsequent $\ell_j^\flat \in A_L^\flat$, except $r_1^\flat$ which only appears once. This is in perfect analogy to the previous case and can be seen in figure \ref{fig:base_terms_causal_diagrams}, where all vertices have valency two, except the top left and bottom right vertex, which have valency one.\\
Condition $3)$ prohibits us from combining $a_k \in A \setminus A^\flat$ with an $\ell_j^\flat \in A_L^\flat$ such that $a_k$ would be smaller or larger than an $r_i^\flat$ accompanying $\ell_j^\flat$. 
The only possibility is thus to combine $a_k$ with the left index that comes in combination with $r_t^\flat$ and $r_{t+1}^\flat$. 
Then $a_k$ is neither the smallest nor the largest index and hence we avoid incompatibilities with the remaining factors of the term.
Consequently, we can multiply the factor
\begin{align}
    \label{eq:rec_ak_right}
    (\ell_{m-t+1}^\flat,a_k)
\end{align}
with expression \eqref{eq:base_term_recursion}.\\
The addition of an index $a_k \in A_R$ to the base term is illustrated in figure \ref{fig:adding_right_vertex}. The black lines build again the base term and each red line connects it to a new vertex, in analogy to figure \ref{fig:adding_left_vertex}.\\
For $t=0$ the term in equation $\eqref{eq:rec_ak_right}$ is not defined because $m+1 > m = N_L^\flat$. Note that in contrast to the case $s=0$, the case $t=0$ can in principle occur since there is no restriction for $r_1^\flat$.
Looking at figure \ref{fig:adding_right_vertex}, or considering condition 3), it is clear that for $t=0$ we can only add 
\begin{align}
    \label{eq:rec_ak_right_t=0}
    (\ell_{m}^\flat,a_k).
\end{align}
However, if we consider the base term with $r_1^\flat$ replaced with $a_k$, we obtain the same term for $t=1$ as we would for the previous base term for $t=0$. Therefore, excluding the case $t=0$ prevents not only the introduction of a special case in equation \eqref{eq:rec_ak_right}, but also double counting of terms generated by the algorithm.\\
This point is illustrated in figure \ref{fig:t=0_illustration} for the $5$-point function. One can see that the case $t=0$ for the diagram on the left-hand side yields the same term as the case $t=1$ on the right-hand side.
It is now natural to ask whether we can have double counting also for $t \neq 0$, \textit{e.g.} by exchanging a line ending on $r_j^\flat$ with one ending on $a_k$ for $j \neq 1$. The short answer is no because $r_1^\flat$ corresponds to the only right vertex with valency one, while all others have valency two. The diagram would be disconnected and hence it would not correspond to a valid term. The topic of uniqueness of terms will be discussed in more detail in section \ref{sec:uniqueness_of_terms}.\\
From this analysis we know that if $t \neq 0$, we have two terms that are compatible with the base term of our recursion and incompatible with each other, so we can directly multiply the sum of these two terms,
\begin{align}
    \label{eq:ak_sum_recursion}
    (a_k,r_{m-s+1}^\flat) + (\ell_{m-t+1}^\flat,a_k)
\end{align}
to \eqref{eq:base_term_recursion}. 
For $t=0$ we only multiply by the term in eq. \eqref{eq:rec_ak_left}, which is the one coming from adding a left index.

\subsubsection*{Adding multiple vertices}
\begin{figure}[ht!]
    \centering
    \includegraphics[width=0.9\textwidth]{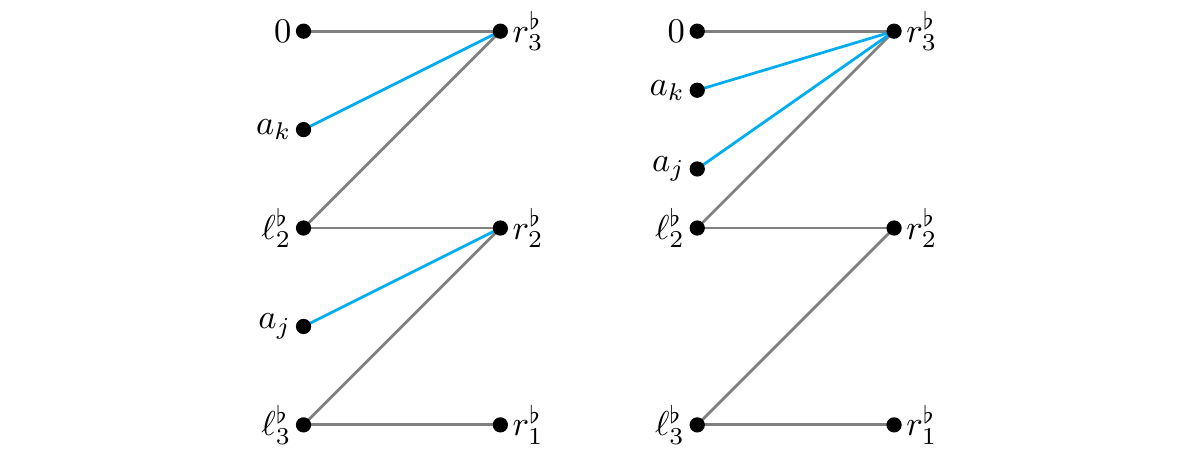}
    \caption{\small Causal diagram for adding two left vertices to base terms with $a_k<a_j$ for the $8$-point function.}
    \label{fig:adding_two_vertices}
\end{figure}

\begin{figure}[ht!]
    \centering
    \includegraphics[width=0.9\textwidth]{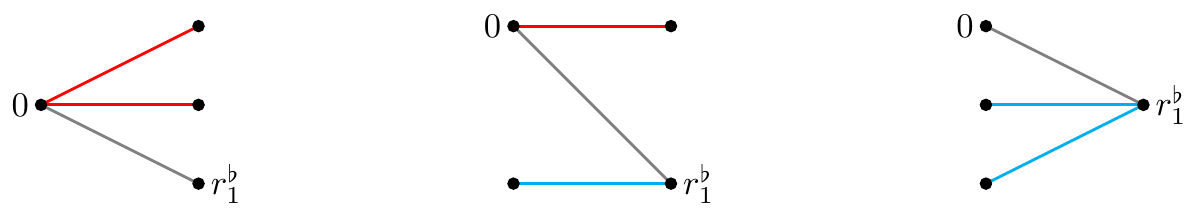}
    \caption{\small Topologies generated by $m=1$ base terms for $N=4$.}
    \label{fig:missing_term}
\end{figure}

As a next step let us assume that we have already multiplied our base term by the above expressions for some $a_j \in A$. Let $a_k \in A \setminus A^\flat$, $a_k \neq a_j$ be another index that we want to include in our term. The rules of multiplying terms \eqref{eq:rec_ak_left} or \eqref{eq:ak_sum_recursion} are the same as before adding the index $a_j$ to the base term, as we will now show.\\
Consider first the case of adding a left vertex, $a_j \in A_L$. We have to consider two cases for multiplying a new factor including $a_k$ to this term. The first case is given by $\ell_s^\flat < a_k < \ell_{s+1}^\flat$, \textit{i.e.}, the index $a_j$ is either smaller than $\ell_s^\flat$ or larger than $\ell_{s+1}^\flat$ and has thus no effect on the multiplication. Consequently, nothing changes in this case with respect to before. This case is shown on the left-hand side of figure \ref{fig:adding_two_vertices}.\\
As a next case we have $\ell_s^\flat < a_j < a_k < \ell_{s+1}^\flat$ or $\ell_s^\flat < a_k < a_j < \ell_{s+1}^\flat$. In the first case, which is shown on the right of figure \ref{fig:adding_two_vertices}, the relevant part of the term to which we want to add the index $a_k$ is
\begin{align*}
    (\ell_s^\flat,r_{m-s+1}^\flat)(\ell_s^\flat,r_{m-s+2}^\flat)(a_j,r_{m-s+1}^\flat)(\ell_{s+1}^\flat,r_{m-s}^\flat)(\ell_{s+1}^\flat,r_{m-s+1}^\flat).
\end{align*}
This term was constructed so that the partner of $a_j$ (\textit{i.e.}, the index accompanying $a_j$), which is $r_{m-s+1}^\flat$, is the smallest partner of $\ell_s^\flat$ and the largest partner of $\ell_{s+1}^\flat$. The partner of $a_k$ must consequently be $r_{m-s+1}^\flat$, since it is the smallest partner of $a_j$ and the largest partner of $\ell_{s+1}^\flat$.\\
The case $\ell_s^\flat < a_j < a_k < \ell_{s+1}^\flat$ can be discussed analogously. The smallest partner of $\ell_s^\flat$ is still $r_{m-s+1}^\flat$, which is also the largest partner of $a_j$.
The discussion works analogously for adding right indices. \\ \\
This can be generalised to arbitrarily many indices $a_j$ that are added to the base term. 
We have thus shown that the rules for multiplying terms \eqref{eq:rec_ak_left} and \eqref{eq:ak_sum_recursion} apply not only for base terms but for all terms that can be constructed with these rules.\\ \\
Algorithm \ref{alg:algorithm_first_step} shows the basic principle  of the generation of causal terms. Here $\mathcal{B}$ denotes the set of base terms and $\mathcal{R}_B$ denotes the array which stores the results of the multiplications for a base term $B \in \mathcal{B}$. The employment of a different $\mathcal{R}_B$ for each base term is necessary in order to avoid multiplications of expressions \eqref{eq:rec_ak_left} or \eqref{eq:ak_sum_recursion} to terms that correspond to a different base term $B'$ that is not compatible with the former terms. The indices $s$ and $t$ are defined as before.
\begin{algorithm}
\caption{\small Generation of causal terms}
\label{alg:algorithm_first_step}
\begin{algorithmic}[1]
\For{$B \in \mathcal{B}$}
    \State $\mathcal{R}_B$ stores $B$
    \For{$a_k \in A \setminus A^\flat$}
        \If{$t \neq 0$}
            \For{$R \in \mathcal{R}_B$}
                \State $\mathcal{R}_B$ stores $R \cdot \left[(a_k,r_{m-s+1}^\flat) + (\ell_{m-t+1}^\flat,a_k)\right]$
            \EndFor
        \Else
            \For{$R \in \mathcal{R}_B$}
                \State $\mathcal{R}_B$ stores $R \cdot (a_k,r_{m-s+1}^\flat)$
            \EndFor
        \EndIf
    \EndFor
\EndFor
\end{algorithmic}
\end{algorithm}
The first loop in algorithm \ref{alg:algorithm_first_step} goes over all base terms. As explained above, we employ for each base term an individual $\mathcal{R}_B$ to save all results constructed from the base term $B$. Then we have to loop over all $a_k \in A$, that are not already elements of $A^\flat$, since we must introduce a new index to the term. The distinction between $t = 0$ and $t \neq 0$ is necessary to avoid double counting. Finally, we multiply not only the base term, but also all terms that are constructed from it, \textit{i.e.} all terms in $\mathcal{R}_B$, by either expression \eqref{eq:rec_ak_left} or \eqref{eq:ak_sum_recursion}. \\ \\
As a final remark of this section let us explicitly highlight topologies we generate from base terms with $m=1$ for $N=4$ in figure \ref{fig:missing_term}.
Note that there is no base term generated for $m=2$, which would correspond to an $s=0$. This is due to the fact that we sort vertices from low to high on the left side and there exists no vertex with index lower than $0$. 
Therefore we will discuss the generation of base terms to complete our algorithm in the next section.

\subsubsection{Proof of the recursion start}
\label{sec:recursion}
The last missing piece for the generation of causal terms is the generation of the base terms. We will show in this section that the base terms are indeed of the form given in eq. \eqref{eq:base_term_recursion}.\\
The very first step of the algorithm is to generate all E-surfaces, which will be stored in a matrix so that they are computed only once in each step. The E-surfaces which contain the $0$-index are now used as base terms. We only use those E-surfaces since all terms without the $0$ as index do not contribute to the amplitude. Note that the E-surfaces are of the form of eq. \eqref{eq:base_term_recursion} since the empty product is defined to be one.\\
Now algorithm \ref{alg:algorithm_first_step} generates unique causal terms where each factor shares exactly one index with the base term. As becomes clear from the argumentation in section \ref{sec:generation_of_causal_terms}, we obtain all of these terms. Consequently, we only miss the terms that contain at least one factor that does not have an index in common with the respective base term. The lowest-point terms with this property will be our new base terms for the next step of the recursion.\\
We find that 3-point terms do not have this property since they consist only of two factors. Otherwise they would have $4$ distinct indices but they must have exactly $3$. The new base terms are thus $4$-point terms of the form
\begin{align*}
    (0,a_i)(a_k,a_j)(a_k,a_i),
\end{align*}
where $j < i$, $k \neq i$ and $k \neq j$. These terms are again of the form of eq. \eqref{eq:base_term_recursion}.\\
After rerunning algorithm \ref{alg:algorithm_first_step} with the new base terms, we now obtain all causal terms except the ones with one or more factors that share no index with the respective base term, or alternatively, with at least two factors that have no index in common with the original E-surface used to generate the new base term. The same argument that applied to the $3$-point terms applies now to the $5$-point terms, so that the lowest-order terms with the desired property are $6$-point terms of the form of eq. \eqref{eq:base_term_recursion}.\\
This goes on until all terms are generated.

\subsubsection{Uniqueness of terms}
\label{sec:uniqueness_of_terms}

It is important to be aware of potential double countings of terms generated by algorithm \ref{alg:algorithm_first_step}. We will now show that each term that is generated by algorithm \ref{alg:algorithm_first_step} is unique. For a single base term, this is easy to see. Since we work with ordered sets we have a unique order of multiplications of terms containing new indices $a_k \in A \setminus A^\flat$. A term containing $a_j$ and $a_k$ with $j < k$ can only be generated by first multiplying the base term with the term containing $a_j$ and then multiplying it with the term containing $a_k$, and not the other way around. Consequently, the way a term is generated is unique so that we have no double counting in this case.\\
Next we have to show that starting with a base term $B$, algorithm \ref{alg:algorithm_first_step} does not generate terms that can also be generated by starting with a different base term $B'$. If we denote with $C$ the product of terms that is multiplied with the base term, we want to show that
\begin{align}
    \label{eq:rec_uniqueness_of_terms_proof}
    B \cdot C \neq B' \cdot C'
\end{align}
for any base terms $B$ and $B'$ and any valid products of causal terms $C$ and $C'$. In other words, we want to show that there is no subexpression in $B \cdot C$ that can be identified with another base term $B'$. To prove this we make use of the fact that in each base term, all left indices $\ell^\flat \in A_L^\flat$ and all right indices $r^\flat \in A_R^\flat$ appear exactly twice, except $\ell_1^\flat = 0$ and $r_1^\flat$. This can be directly deduced from eq. \eqref{eq:base_term_recursion}.\\
Multiplication with the expression in \eqref{eq:rec_ak_left} yields a new left index that appears only once because the new index has only one compatible right index $r \in A_R^\flat$. We can thus conclude that the base term $B'$ must have the same set $A_L^\flat$ as $B$. The same argument holds for multiplication with the factor in eq. \eqref{eq:rec_ak_right} except that the new index $a_k$ can be smaller than $r_1^\flat$. In this case, however, the term in eq. \eqref{eq:rec_ak_right} does not exist, as argued below eq. \eqref{eq:rec_ak_right}. Thus we find that the base term $B'$ must also have the same set $A_R^\flat$ as $B$. Together with the fact that all base terms possess the same topology, \textit{cf.} figure \ref{fig:base_terms_causal_diagrams}, this shows that eq. \eqref{eq:rec_uniqueness_of_terms_proof} is true and thus concludes the proof of uniqueness.

\subsubsection{Scaling behaviour}
\begin{figure}[ht]
    \centering
    \includegraphics[width=0.49\textwidth]{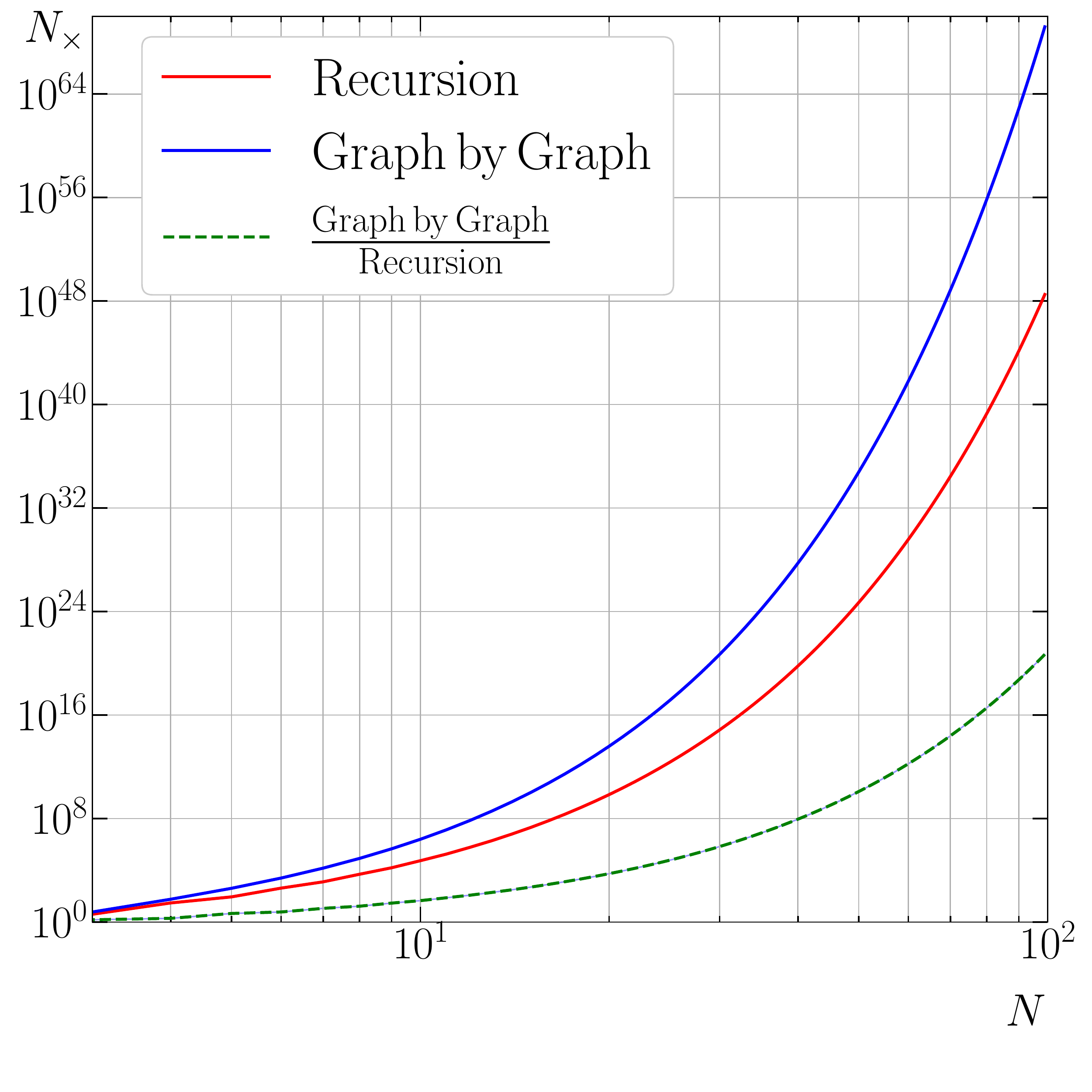}
    \includegraphics[width=0.49\textwidth]{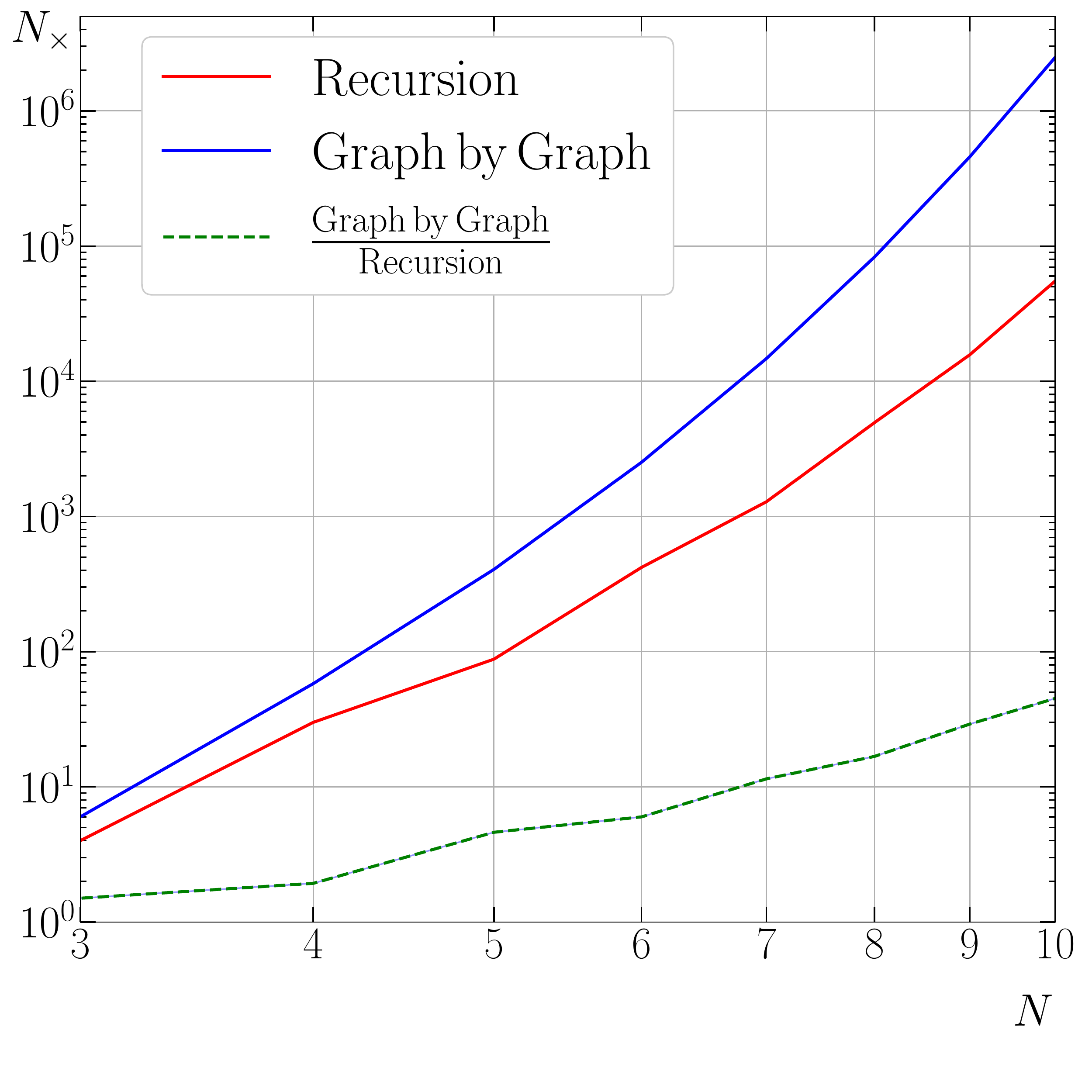}
    \caption{\small Number of multiplications for one cyclic order for the recursion (red) compared to the graph-by-graph approach (blue) with respect to the number of external legs (left: $3 \leq N \leq 100$; right: $3 \leq N \leq 10$). The ratio of these two scalings is shown as a dashed green line in both plots.}
    \label{fig:scaling}
\end{figure}

In this section we want to investigate the scaling behaviour of the recursion compared to the na\"{i}ve graph-by-graph approach, where one sums up all contributions from all relevant Feynman graphs. We assume that the amplitude is structured as described in section \ref{sec:structre_of_the_amplitude} in both approaches. Consequently, the factor $(N-1)!/2$ appears in both cases, so we only consider scaling for one cyclic order. Moreover, we assume that the E-surfaces are pre-computed and count only the multiplications of E-surfaces which makes it much easier to compute the scaling behaviour. The cost of the computation of E-surfaces is negligible compared to the rest of the amplitude.\\
To obtain the behaviour for the graph-by-graph approach, we recall that each $n$-point function consists of $\binom{2(n-1)}{n-1}$ causal terms, \textit{cf.} eq. \eqref{eq:number_of_causal_terms_diagram}. Furthermore, as explained in section \ref{sec:structre_of_the_amplitude}, there are $\binom{N-1}{n-1}$ different $n$-point functions in an $N$-point amplitude. Combining these two facts yields the total number of causal terms for one cyclic order in an $N$-point amplitude,
\begin{align*}
    N_\text{terms}^\text{graph} = \sum_{j=0}^{N-2} \binom{N-1}{j+1} \binom{2(j+1)}{j+1}.
\end{align*}
Since we want to count the number of multiplications, we have to multiply each term in the sum by $j$,
\begin{align}
    \label{eq:scaling_gbg}
    N_\times^\text{graph} = \sum_{j=1}^{N-2} j \binom{N-1}{j+1} \binom{2(j+1)}{j+1},
\end{align}
due to the fact that each causal $n$-point term is a product of $(n-1)$ E-surfaces.\\
To obtain the number of multiplications of our algorithm, we proceed in two steps. First we compute the number of terms that algorithm \ref{alg:algorithm_first_step} generates, and then proceed with the number of base terms.\\
We can multiply a base term corresponding to an $n$-point function, \textit{i.e.}, $|A^\flat| = n$, with maximal $(N-n)$ terms so that all of these terms contribute to the $N$-point amplitude. This leads to $\binom{N-(n-1)}{2}$ multiplications in algorithm~\ref{alg:algorithm_first_step} for the construction of all possible terms resulting from one base term. Recall that for all base terms $n$ is an even integer, $n = 2m$, where $m = 1, \dots, \lfloor \frac{N-1}{2} \rfloor$, and that $m = N_L^\flat = N_R^\flat$. Thus, there are $\binom{2m}{m}$ different base terms for each $n$-point function. Together with the number of $n$-point functions within a cyclic permutation of external legs of an $N$-point amplitude we obtain,
\begin{align}
    \sum_{j=1}^{\lfloor \frac{N-1}{2} \rfloor} \binom{N-1}{2j-1} \binom{2j}{j} \binom{N-(2j-1)}{2}.
\end{align}
The second contribution comes from the construction of the base terms, which involves two multiplications per term. This yields
\begin{align}
    2\sum_{j=1}^{\lfloor \frac{N-2}{2} \rfloor} \binom{N-1}{2j-1} \binom{2j}{j}.
\end{align}
The full expression reads,
\begin{align}
    \label{eq:scaling_recursion}
    N_\times^\text{rec} = \begin{cases}
                            \sum_{j=1}^{\frac{N-2}{2}} \binom{N-1}{2j-1} \binom{2j}{j} \left[ \binom{N-(2j-1)}{2} + 2\right], & N \, \text{is even},\\
                            \sum_{j=1}^{\frac{N-3}{2}} \binom{N-1}{2j-1} \binom{2j}{j} \left[ \binom{N-(2j-1)}{2} + 2\right] + \binom{N-1}{N-2} \binom{N-1}{(N-1)/2}, & N \, \text{is odd}.
                        \end{cases}
\end{align}
The number of multiplications depends on whether $N$ is even or odd, which can be explained by recalling that the base terms exist only for even $n$, as mentioned above.\\
The scaling with respect to the number of external legs is shown as double logarithmic plot in figure \ref{fig:scaling}. It compares the scaling of the na\"ive graph-by-graph approach (blue) to the scaling of the recursion (red) presented in this section. The ratio of both scalings is shown in both plots as a dashed green line. 
One can see in the plot on the left hand side, where $3\leq N \leq 100$, that both approaches scale exponentially due to the exponential growth of Feynman diagrams within the amplitude and the additional exponential growth of causal terms per Feynman diagram. However, the recursion scales better than the graph-by-graph approach and the ratio between the two grows exponentially.\\
Furthermore, the scaling of the recursion is not smooth but depends on whether we go from an even number of external legs to an uneven one or the other way around, which can be seen in the plot on the right-hand side, where the horizontal axis is zoomed in to the practically relevant region $3 \leq N \leq 10$. This behaviour is not surprising if one recalls equation \eqref{eq:scaling_recursion}, and can be traced back to the base terms which appear only for even $n$.

\subsection{Tree currents}
\label{sec:rec_tree_currents}
\begin{figure}[ht]
    \centering
    \includegraphics[width=0.6\textwidth]{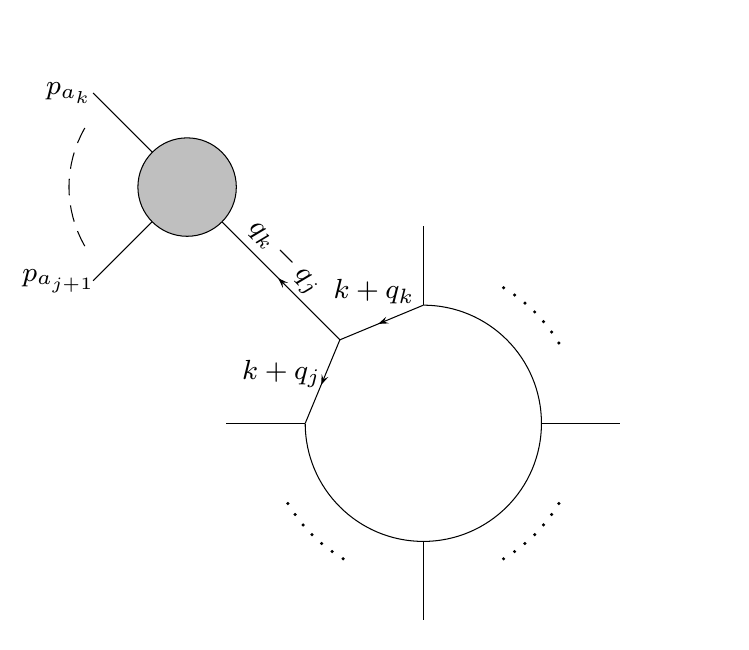}
    \caption{\small A one-loop $N$-point scalar diagram with a tree carrying the momenta $p_{a_{j+1}}, \dots, p_{a_k}$, where $k > j$. The tree is denoted by a grey circle. The arrows indicate the direction of the momentum flow.}
    \label{fig:tree_currents}
\end{figure}
Up to now we have only talked about the causal terms. In order to compute a complete scattering amplitude, we also need to take the trees into account that are part of the one-loop Feynman diagrams. To do so, we note that there is a one-to-one correspondence between products of tree currents and causal terms.\\
Recall that the indices of a causal term correspond to momenta flowing through the propagators. Specifically for an index $a_k$ we have the momentum $q_k$, defined as
\begin{align}
    \label{eq:correspondence_index_momenta}
    q_k \coloneqq \sum_{i=1}^k p_{a_i}.
\end{align}
This is just a generalisation of the definition in equation \eqref{eq:simple_lin_comb_ext_moms} to arbitrary permutations of the external legs.\\
Let us consider a term with subsequent indices $a_j$ and $a_k$ with $j < k$. From this we can infer that the term corresponds to a Feynman graph with a vertex connecting the propagators with momenta $q_j$ and $q_k$. This is illustrated in figure \ref{fig:tree_currents}. Since throughout this paper we work in $\phi^3$ theory, we know that there must be a third edge to this vertex that is part of a tree. The external legs of this tree are dictated by momentum conservation because the third propagator must carry the momentum
\begin{align}
    \label{eq:tree_momentum}
    q_k - q_j = \sum_{i=j+1}^k p_{a_i}.
\end{align}
However, it does not determine how the legs are ordered, and, in fact, all permutations of these external legs contribute to the amplitude. Thus, we can directly multiply the causal term with a tree current, which can be generated by employing the Berends-Giele recursion \cite{Berends:1987me}.\\
We can conclude that all tree currents for a causal term are determined by its set of indices, $A$, by iterating through the set and deducing the external momenta from the gap between two neighbouring indices, as shown in equation \eqref{eq:tree_momentum}.

\subsection{Symmetry factors}
\label{sec:recurrence_relations_sym_facs}
The last missing piece of our algorithm are the symmetry factors that take into account multiple appearances of terms in different cyclic orderings. Before we provide a combinatorial derivation, we briefly comment on the intuition in which contexts they arise.\\

\subsubsection{Intuitive description}
\label{sec:intuitive_description}
\begin{figure}[t]
    \centering
    \includegraphics[width=0.6\textwidth]{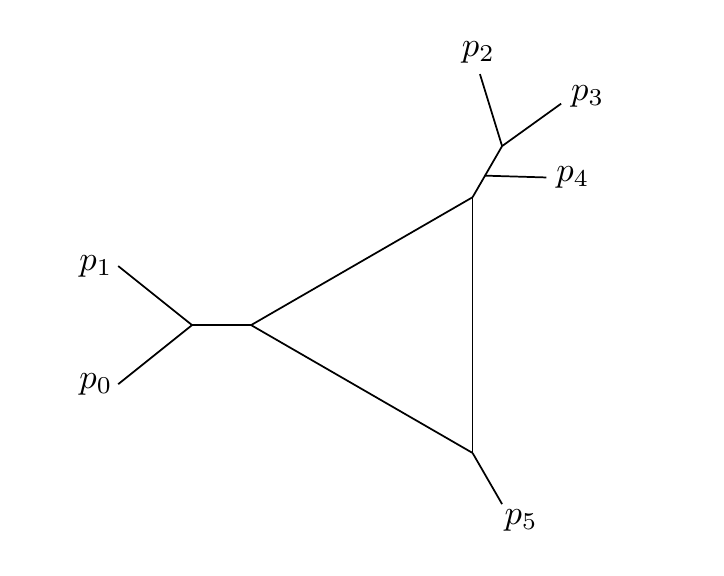}
    \caption{\small A 6-point triangle diagram with non-trivial trees attached to some of the legs exiting the loop.}
    \label{fig:tree_perms}
\end{figure}
The symmetry factors are closely related to the tree factors multiplying the loop integrands.
Let us consider a one-loop $n$-point function which can have non-trivial trees connected to the loop propagators, like the one shown in figure \ref{fig:tree_perms}.
There are three different sources causing double counting.\\
The first one concerns trees like the one with legs $p_{2}$ and $p_{3}$ in figure \ref{fig:tree_perms} where we are free to swap these two legs without altering the numerical value associated to this diagram. These two diagrams are generated in two different cyclic orders, so we have to account for that by introducing a symmetry factor. Note that the tree with legs $p_{0}$ and $p_{1}$ does not introduce a symmetry factor because we fix the leg $p_{0}$ in order to exclude the cyclic permutations of the legs.\\
The second source is related to the fact that we multiply the loop integrands by tree currents instead of individual trees. The tree currents contain trees corresponding to all permutations of external legs that yield a different Feynman diagram with a different numerical value. As an example, consider the tree consisting of the three external legs $p_{2}$, $p_{3}$ and $p_{4}$ of the diagram in figure \ref{fig:tree_perms}. Exchanging $p_{2}$ and $p_{4}$ yields a different tree belonging to the same current as the original tree. Consequently, we encounter this diagram also in a different cyclic order. The first two sources taken together yield a symmetry factor of $1/j!$ for each tree with $j$ external legs if none of these legs is $p_0$, and $1/(j-1)!$ otherwise.\\
The third source for the double counting of diagrams comes from the reflection symmetry of diagrams, or equivalently the fact, that we can choose a different direction for the routing of the loop momentum $k$. This leads to different propagator momenta and thus contributes to a different cyclic order. Therefore we obtain an additional factor of $1/2$ for our symmetry factor.\\
Note that changing the direction of the momentum flow leads to the same propagator momenta in the case of bubble diagrams, since we always keep $p_0$ fixed. Consequently, the two propagator momenta are in both cases given by $k$ and $k + q$, where $q$ is the linear combination of external momenta of the tree which has not $p_0$ attached to it. Thus, the factor of $1/2$ seems to be unnecessary in the case of bubbles. However, bubbles naturally come with a symmetry factor of $1/2$, so that we obtain the same expression for symmetry factors for all $n$-point functions within the $N$-point amplitude.

\subsubsection{Derivation of the symmetry factor}
As becomes clear from the discussion above, the symmetry factors appear on the level of Feynman diagrams, \textit{i.e.}, all causal terms corresponding to the same Feynman diagram obtain the same symmetry factor. More explicitly, the loop integrand determines the possible tree currents that can be multiplied to it and therefore it determines also the symmetry factor. The loop integrand of an $n$-point function can be specified by the set $A_n$, which contains $n$ indices corresponding to the $n$ loop propagators. We write the subscript $n$ to distinguish it from the set $A$, which contains all $N$ indices.\\
Let us now turn to the derivation of the symmetry factors. While in the previous sections we treated the cyclic orders as self-contained separated pieces so that everything applied equally to all cyclic orders, we now loosen this structure and deal with the whole amplitude. This introduces a new validity constraint for causal terms that was guaranteed in the previous case and which will be derived in the following. The new constraint relies on the fact that an index corresponds to a specific linear combination of external momenta, as indicated by equation \eqref{eq:correspondence_index_momenta}.
In a one-loop diagram, the momentum flowing through a propagator is a linear combination of momenta of external particles that must be a subset or a superset of the external momenta corresponding to a neighbouring propagator. This is fulfilled by construction within an individual cyclic order because of equation \eqref{eq:correspondence_index_momenta}. In this section we need to enforce this constraint.\\
We want to derive the symmetry factors by counting how often a certain combination of indices appears in the amplitude, based on the symmetry of the cyclic orderings.\\
Furthermore, we can assign to each index the number of external momenta participating in the sum of equation \eqref{eq:correspondence_index_momenta}. For an $N$-point function within a one-loop $N$-point amplitude where we eliminate the largest index by means of momentum conservation we have for each number $j$ of external particles exactly one index $a_{j+1} \in A$, where $0 \leq j \leq N-1$. Note that $j$ starts at zero because $a_1 = 0$ and therefore the momentum of the corresponding propagator consists entirely of the loop momentum $k$. Hence, no external momentum flows through this propagator and $j$ must be zero.\\
Our strategy is to first compute the number of times that an individual index appears in the amplitude. Then we want to find out the same for a specific valid combination of two indices before generalising this to $n$ indices, where $n \leq N-1$.\\
There is exactly one index for each number of external momenta in a set $A$, and considering all permutations of the momenta we have $\binom{N-1}{j}$ different linear combinations of $j$ external momenta in the amplitude. In total, we have $(N-1)!/2$ cyclic permutations of external legs, so that the number of times a specific index corresponding to a linear combination of $j$ external momenta appears in the amplitude is given by
\begin{align}
    \label{eq:j_moms_tot}
    \frac{(N-1)!}{2\binom{N-1}{j}} = \frac{(N-1-j)! \, j!}{2}.
\end{align}
Next, we want to compute the number of times a specific combination of $i$ and $j$ momenta occurs, where $i > j$. Recall that the set of $j$ external momenta has to be a subset of the set of $i$ external momenta.
The number of combinations of $i$ different external momenta is given by the binomial coefficient  $\binom{N-1}{i}$. 
Each of these combinations contains $\binom{i}{j}$ combinations of $j$ different external momenta, so that $\binom{N-1}{i} \cdot \binom{i}{j}$ is the total number of combinations of $j$ external momenta that occur in the $\binom{N-1}{i}$ combinations of $i$ momenta. These combinations are not all distinct. We have $\binom{N-1}{j}$ distinct combinations of $j$ external momenta, so to obtain the number of times a specific combination of $j$ external momenta occurs in the $\binom{N-1}{i} \cdot \binom{i}{j}$ combinations, we have to divide by $\binom{N-1}{j}$. This yields
\begin{align}
    \label{eq:num_sub_comb_j}
    \binom{N-1}{i} \frac{\binom{i}{j}}{\binom{N-1}{j}} = \binom{N-1-j}{N-1-i}.
\end{align}
Now we can determine the number of times a specific combination of $i$ and $j$ momenta occurs, as the number of times the index $j$ occurs, (eq. \eqref{eq:j_moms_tot}) divided by the number of different combinations $i$ that go along with the same combination $j$ (eq. \eqref{eq:num_sub_comb_j}),
\begin{align}
    \label{eq:num_specific_comb}
   \frac{(N-1)!}{2 \binom{N-1}{j} \binom{N-1-j}{N-1-i}} = \frac{(N-1-j)! \, j!}{2 \binom{N-1-j}{N-1-i}} = \frac{(N-1-i)! \, (i-j)! \, j!}{2}.
\end{align}
This can be generalised to finding the number of combinations of $n \leq N$ indices by using the set $A_n$, consisting of the $n$ indices. Each index must correspond to a different number of external particles. We can generalise equation \eqref{eq:num_specific_comb} to
\begin{align*}
    \frac{(N-1)!}{2 \prod_{i = 1}^n \binom{N-1-j_{i-1}}{N-1-j_i}} = \left[\prod_{i=1}^n (j_i - j_{i-1})!\right] \,  \frac{(N-1-j_n)!}{2},
\end{align*}
where $j_i \in A_n$ for $0 < i \leq n$ and $j_0 \equiv 0$.\\
The remaining part of the symmetry factor consists of a factor of $2$ for each term, which originates from the fact that we can reverse the momentum flow, which will always yield a different combination of indices for $n \geq 3$ and thus appears in the calculation. As discussed in section \ref{sec:intuitive_description}, the factor of $2$ is also true for $n=2$, but the reason for the appearance of this factor is different. The symmetry factor reads,
\begin{align*}
    S = \left(\left[\prod_{i=1}^n (j_i - j_{i-1})!\right] \,  (N-1-j_n)!\right)^{-1},
\end{align*}
which is true for all $n \geq 2$.

\section{UV renormalisation}
\label{sec:UV_renormalisation}

The $N$-point one-loop amplitude in scalar $\phi^3$-theory requires
ultraviolet renormalisation. In this section we give the details how this is done within a numerical approach based on causal loop-tree duality.
Parts of the discussion follow ref.~\cite{Baumeister:2019rmh}.

We work in dimensional regularisation with $D=4-2\epsilon$ and an arbitrary scale $\mu$ with mass dimension one.
To be consistent with existing literature we define scalar one-loop integrals in this section as
\begin{align}
    \begin{split}
    A_0(m^2) &= e^{\epsilon \gamma_E}\mu^{2\epsilon} \int\frac{d^Dk}{i \pi^{D/2}} \frac{1}{k^2-m^2},\\
	B_0(p^2,m_1^2,m_2^2) &= e^{\epsilon \gamma_E}\mu^{2\epsilon} \int\frac{d^Dk}{i \pi^{D/2}} \frac{1}{(k^2-m_1^2)((k-p)^2-m_2^2)},\\
	C_0(p_1^2,p_2^2,p_3^2,m_1^2,m_2^2,m_3^2) &=\\ e^{\epsilon \gamma_E}\mu^{2\epsilon} \int\frac{d^Dk}{i \pi^{D/2}} &\frac{1}{(k^2-m_1^2)((k-p_1)^2-m_2^2)((k-p_1-p_2)^2-m_3^2)},
    \end{split}
\end{align}
with $p_3^2=(p_1+p_2)^2$.

\subsection{Scalar \texorpdfstring{$\phi^3$}{phi3} renormalisation}

We work with a scalar massive $\phi^3$ theory with the Lagrangian 
\begin{align}
    \mathcal{L}_0 = \frac{1}{2}(\partial_{\mu}\phi_0)(\partial^{\mu}\phi_0) - \frac{1}{2} m_0^2 \phi_0^2 + \frac{1}{3!} \lambda_0^{(D)}\phi_0^3,
\end{align}
where the subscript $0$ indicates bare quantities. 
These are related to the renormalised quantities by the relations
\begin{align}
    \phi_0 = Z_{\phi}^{\frac{1}{2}} \phi, 
    \;\;\;
    \lambda_0^{(D)} = Z_{\lambda}\lambda^{(D)}, 
    \;\;\;
    m_0 = Z_m m.
\end{align}
We set the coupling constant to 
\begin{align}
    \begin{split}
        \lambda^{(D)} = \mu^{\epsilon}S_{\epsilon}^{-\frac{1}{2}} \lambda,
    \end{split}
\end{align}
where $S_{\epsilon} = (4\pi)^{\epsilon}\text{exp}(-\epsilon \gamma_E)$.
The arbitrary scale $\mu$ is introduced to keep the mass dimension of the renormalised coupling $\lambda$ equal to one.
The renormalised Lagrangian is
\begin{align}
    \mathcal{L} = \frac{1}{2}(\partial_{\mu}\phi)(\partial^{\mu}\phi) - \frac{1}{2} m^2 \phi^2 + \frac{1}{3!} \lambda^{(D)}\phi^3 + \mathcal{L}_{\text{CT}} 
\end{align}
with
\begin{align}
    \mathcal{L}_{\text{CT}} = -\frac{1}{2}(Z_{\phi}-1)\phi \Box \phi - \frac{1}{2}(Z_{\text{m}}^2Z_{\phi}-1) m^2 \phi^2 + \frac{1}{3!}(Z_{\lambda}Z_{\phi}^{3/2}-1) \lambda^{(D)}\phi^3.
\end{align}
The Feynman rules for this theory are 
\begin{align}
\begin{split}
    \includegraphics[width=0.9\textwidth]{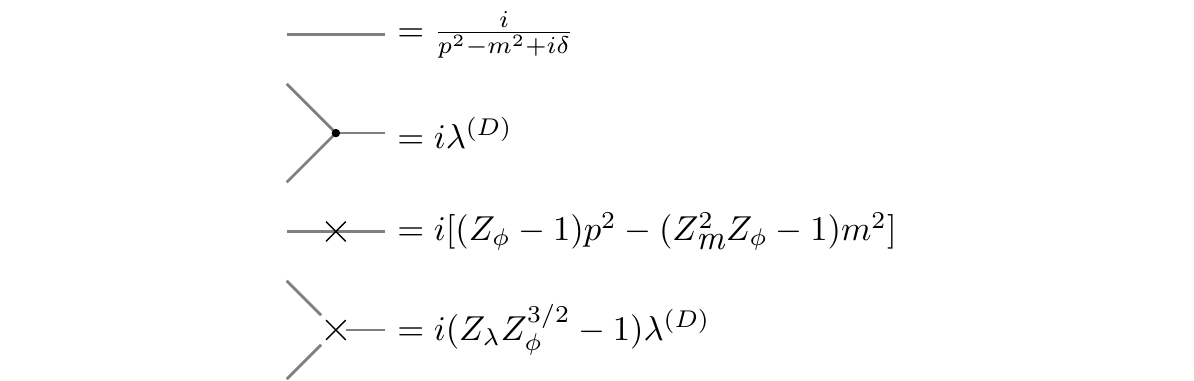}
\end{split}   
\end{align}
The perturbative expansion of the renormalisation constants is 
\begin{align}
    Z_a = 1 + \sum_{n=1}^{\infty} Z_a^{(n)}\left(\frac{\lambda^2}{(4\pi)^2}\right)^n, \;\;\; a \in \{\phi,m,\lambda\},
\end{align}
 and we keep the first order terms for our next-to-leading-order calculation. 

Two popular renormalisation schemes are the on-shell scheme and the $\overline{\text{MS}}$-scheme.

\subsubsection*{On-shell scheme}
In the on-shell scheme we set the $3$-point amplitude with $p_1^2=p_2^2=p_3^2=m^2$ equal to $i\lambda$ to fix $Z_\lambda^{(1)}$.
We find the following values for the renormalisation constants
\begin{align}
    \begin{split}
            Z_\phi^{(1)} &= \frac{2-\epsilon}{6m^2}B_0(m^2,m^2,m^2) - \frac{1-\epsilon}{3m^4} A_0(m^2),\\
            Z_m^{(1)} &= \frac{1}{4m^2} B_0(m^2,m^2,m^2),\\
            Z_\lambda^{(1)} &= -\frac{3}{2}Z_\phi^{(1)} + C_0(m^2,m^2,m^2,m^2,m^2,m^2).
    \end{split}
\end{align}

\subsubsection*{\texorpdfstring{$\overline{\text{MS}}$}{MS}-scheme}
In the $\overline{\text{MS}}$-scheme we find
\begin{align}
    \begin{split}
            Z_\phi^{(1)} &= 0\\
            Z_m^{(1)} 
            &= \frac{1}{4m^2}\left(\frac{1}{\epsilon}-\gamma_E+\ln{4\pi}\right)\\
            Z_\lambda^{(1)} &= 0
    \end{split}
\end{align}
for the renormalisation constants.

\subsection{Local UV subtraction terms}
\label{sec:Local_UV_subtraction_terms}
In order to calculate the integrand numerically we use the UV subtraction terms of \cite{Becker:2010ng} and apply loop-tree duality to the expression. 
We choose ultraviolet subtraction terms that have singularities on
\begin{align}
    \label{eq:UV_propagator}
    (k-Q)^2-\mu_{\text{UV}}^2 = 0,
\end{align}
with $Q$ and $\mu_{\text{UV}}$ both real. 
The UV subtraction will be independent of $Q$ and the mass $\mu_{\text{UV}}$ when adding the analytic expression back to the amplitude according to eq. \eqref{eq:subtraction_method} which will be explained later in this section. 
The only UV-divergent diagrams in massive scalar $\phi^3$-theory are bubbles such that we only need to keep the first order in an expansion of the loop propagators around the hypersurface defined by eq.~\eqref{eq:UV_propagator}
\begin{align}
\begin{split}
  \frac{1}{(k-p)^2-m^2} &=\frac{1}{(k-Q)^2-\mu_{\text{UV}}^2} +\mathcal{O}\left(\frac{1}{|k|^3}\right).
\end{split}
\end{align}
The local subtraction term is hence given by
\begin{align}
\begin{split}
    \label{eq:UV_subtraction_term}
    I_{\text{bub}}^{\text{UV}_\text{sub}}(\mu_{\text{UV}}^2) &= \mu^{2\epsilon}\int\frac{d^Dk}{(2\pi)^D}\frac{1}{((k-Q)^2-\mu_{\text{UV}}^2)^{2}}\\
    &= \mu^{2\epsilon}\frac{i}{4}\int\frac{d^{D-1}k}{(2\pi)^{D-1}}\frac{1}{\left( (\Vec{k}-\Vec{Q})^2+\mu_{\text{UV}}^2\right)^{3/2}}\\
    &= \frac{i}{(4\pi)^2}\left(\frac{1}{\epsilon} - \gamma_E + \ln{(4\pi)} - \ln{ \frac{\mu_{\text{UV}}^2}{\mu^2}} + \mathcal{O}(\epsilon)\right),
\end{split}
\end{align}
where we applied loop-tree duality in the second equality. 
This expression matches the $1/\epsilon$ divergence of scalar bubble diagrams. \newline
The relation between the bare and renormalised $n$-point amplitude in $\phi^3$ theory is
\begin{align}
\begin{split}
    \mathscr{A}(p_1,...,p_n,\lambda,m) &= \left(Z_{\phi}^{1/2}\right)^n  \mathscr{A}^{\text{bare}}(p_1,...,p_n,\lambda_{\text{bare}},m_{\text{bare}})\\
    &= \left(Z_{\phi}^{1/2}\right)^n  \mathscr{A}^{\text{bare}}(p_1,...,p_n,Z_{\lambda}\lambda,Z_mm),
\end{split}
\end{align}
and after expanding in the coupling 
\begin{align}
\begin{split}
    \mathscr{A}(p_1,...,p_n,\lambda,m) &= \mathscr{A}^{\text{bare}}(p_1,...,p_n,\lambda,m) + \mathscr{A}^{\text{CT}}(p_1,...,p_n,\lambda,m).
\end{split}
\end{align}
$\mathscr{A}^{\text{CT}}$ contains all the counterterms coming from renormalisation. Even though the sum of the bare and counterterm amplitude is UV finite, each amplitude itself is not. Therefore we introduced local subtraction terms. Adding and subtracting them \cite{Becker:2010ng},
\begin{align}
    \label{eq:subtraction_method}
    \mathscr{A} &= \left(\mathscr{A}^{\text{bare}} - \mathscr{A}^{\text{UV}}\right) + \left(\mathscr{A}^{\text{CT}} + \mathscr{A}^{\text{UV}}\right),
\end{align}
leaves the first bracket locally finite, such that it can be integrated numerically with Monte Carlo methods and the second bracket can be calculated analytically. 
This gives us the freedom to implement any renormalisation scheme: The choice of the renormalisation scheme affects only 
$\mathscr{A}^{\text{CT}}$.

\section{Contour Deformation}
\label{sec:contour_deformation}
Having dealt with UV-divergences, there are only threshold singularities remaining since we work in a massive theory with no IR-divergences. 
It is possible to integrate the threshold singularities numerically using contour deformation as already shown in the literature \cite{Gong:2008ww,Becker:2012aqa,Becker:2012nk,Becker:2012bi,Buchta:2015wna,Capatti:2019edf}. We follow these references. \newline

We deform the $D-1$-dimensional loop momentum to the complex space
\begin{align}
    \label{eq:deformation_direction}
    \Vec{k} = \Vec{k'} + i \lambda\vec{\kappa}(\Vec{k'}),
\end{align}
with $\Vec{k'}$ being real.
$\vec{\kappa}(\Vec{k'})$ sets the deformation direction and $\lambda$ the magnitude.
This deformation introduces a Jacobian, 
\begin{align}
    \left|\frac{\partial \Vec{k}}{\partial \Vec{k'}}\right|,
\end{align}
which we compute numerically. 
Recall that within the recursive implementation of the $N$-point amplitude we define a contour deformation for each cyclic order.
In terms of Feynman graphs, there will be exactly one
$N$-point Feynman graph for a given cyclic order, plus several sub-graphs obtained by pinching
some of the loop propagators.
The deformation for a given $N$-point Feynman graph is also valid for each sub-graph, as each sub-graph contains only a subset of the E-surfaces of its parent graph.
Without loss of generality we fix the cyclic order to $(1,\dots,N)$.
\newline
In this section we make the $i\delta$-prescription explicit by expanding energy factors around small values of $\delta$,
\begin{align}
	&E_i = \color{black}\sqrt{\Vec{k}_i^2+m^2-i\delta}\rightarrow E_i - \frac{i\delta}{2E_i} + \mathcal{O}(\delta^2)
\end{align}
and use the notation of $E_i$ for energy factors without the $i\delta$ for the rest of this section. Further we define $\Vec{k}_i = \Vec{k} + \Vec{q}_i$.
With this we recall, that within causal loop-tree duality we only have E-surfaces with fixed $i\delta$-prescription
\begin{align}
    x_{ij} = \color{black}E_i + E_j - q_{ij}^0 -i\frac{\delta}{2}\left(\frac{1}{E_i} + \frac{1}{E_j} \right),
\end{align}
since in the massive case $E_i>0;\; \forall i$. \newline
We define the set of singular E-surfaces $\mathcal{E}$ for a given set of external momenta $(p_1,...,p_n)$ by the existence condition
\begin{align}
    \label{eq:E_surface_existance}
    q_{ij}^2 \geq 4 m^2.
\end{align}
We check this condition for all $i\neq j$ and put the pair of indices in the set $\mathcal{E}$, if equation \ref{eq:E_surface_existance} is fulfilled. \newline
The contour deformation is defined with technical parameters. 
From an analytical point of view, the contour deformation is independent of these parameters. 
However they still influence the convergence of the integrand and are hence subject to optimisation. 
The parameters are chosen in a way that they work for a generic set of external momenta, even though individually a better performance can be achieved by adjusting the parameters for specific cases.
We used values of \cite{Gong:2008ww,Becker:2012aqa,Becker:2012nk,Capatti:2019edf} as references and expect room for improvement by a proper analysis of these parameters.
We define
\begin{align}
    \sqrt{s} = \underset{\{i,j\}\in \{1,...,N\}, i\neq j}{\text{max}}\left(\sqrt{|(p_i+p_j)^2|}\right)
\end{align}
as a parameter that scales with the external momenta. This is motivated by the centre of mass energy of the system. 
We use $\Vec{k}$ instead of $\Vec{k'}$ for the deformed loop momentum in the following.

\subsection{Deformation direction}
\label{ssec:deformation_direction}
The full deformation direction we use is 
\begin{align}
\label{eq:deformation_direction_vector}
\begin{split}
\vec{\kappa} &= \sum_{\{i,j\} \in \mathcal{E}} g\left(\vec{k}+\frac{1}{2}(\vec{q}_i + \vec{q}_j), q_{i}^0-q_{j}^0, \gamma_g\right) c_{ij}\left(\Vec{k},M_E(q_{i}^0-q_{j}^0)\right)\cdot \; \Vec{b}_{ij}\\
&\phantom{=} + g\left(\vec{k}, M_g, \gamma_g\right) \sum_{a=1}^{N_{\mathrm{soft}}} c_{a}\left(\Vec{k},M_E,\gamma_E\right) \Vec{\kappa}_a
\end{split}
\end{align}
and will be explained in the following. First, the technical parameters are set to 
\begin{align}
\begin{split}
    M_g &= 0.7\sqrt{s}\\
    M_E &= 0.07\\
    \gamma_g &= 0.7\\
    \gamma_E &= 0.008.
\end{split}
\end{align}
Each $N$-point function has a set $\mathcal{E}$ of singular E-surfaces according to the existence condition defined in eq. \eqref{eq:E_surface_existance}. \newline
The first part is a sum of direction vectors $\vec{b}_{ij}$ multiplied by an anti-selection function $c_{ij}$ and a falloff function $g$, over all singular E-surfaces $\mathcal{E}$.
In order to determine the deformation direction $\vec{b}_{ij}$ we apply the deformation of eq. \eqref{eq:deformation_direction} to an E-surface defined in eq. \eqref{eq:E-surface}. 
We denote energy factors, as well as E-surfaces, after the deformation by a prime
\begin{align}
    \label{eq:transformed_energy_factor}
	E_i'&=\sqrt{\Vec{k}_i^2+m^2 + 2i\lambda\Vec{\kappa}\cdot\Vec{k}_i-\lambda^2\Vec{\kappa}^2} \\
	x_{ij}' &= E_i' + E_j' - q_{ij}^0,
\end{align}
and expand around small values of $\lambda$ up to first order in $\lambda$ to make the imaginary part explicit,
\begin{align}
	\label{eq:idelta_prescription}
	\begin{split}
	x_{ij}' &\rightarrow E_i + E_j - q_{ij}^0 + i\lambda \vec{\kappa}\cdot\left(\frac{\vec{k}_i}{E_i} + \frac{\vec{k}_j}{E_j}\right).
	\end{split}
\end{align}
On singular E-surfaces $x_{ij}=0$ the deformation vector $\vec{\kappa}$ should fulfil the $i\delta$-prescription.
The natural choice for one singularity is to choose the vector itself,
\begin{align}
	\begin{split}
	\vec{\kappa} = - \left(\frac{\vec{k}_i}{E_i} + \frac{\vec{k}_j}{E_j}\right).
	\end{split}
\end{align}
However, it turned out that in general the following deformation vector produced better results,
\begin{align}
	\begin{split}
	\vec{\kappa} = - E_j\left(\frac{\vec{k}_i}{E_i} + \frac{\vec{k}_j}{E_j}\right) \equiv \vec{b}_{ij},
	\end{split}
\end{align}
which also has a negative projection on the deformation direction in eq. \eqref{eq:idelta_prescription}. \newline
To define a complete deformation for an $N$-point function we make use of the anti-selection method, setting the deformation to zero whenever it deforms in the wrong direction. 
We define the helper functions, that have been used for contour deformation in \cite{Gong:2008ww},
\begin{align}
    \label{eq:helping_function_theta_one_loop}
    h_{\theta}(t,M_E)&= \frac{t}{t+M_E^2} \theta(t)\\
    \label{eq:helping_function_delta_one_loop}
    h_{\delta}(t,M_E)&= \frac{t^2}{t^2+M_E^2}.
\end{align}
We further define
\begin{align}
\label{eq:cij_helper_function}
c_{ij}\left(\Vec{k},M_E\right) &\equiv \prod_{\{l,m\} \in \mathcal{E}, \{l,m\} \neq \{i,j\} } \text{max}\left[h_{\delta}(x_{lm},M_E),h_{\theta}\left( \color{black}\Vec{b}_{ij}\cdot\Vec{b}_{lm},M_E\right)\right],
\end{align} 
while the product is set to one if $\mathcal{E}$ has only $\{i,j\}$ as element.
Far away from singular surfaces, $c_{ij}\sim 1$, because the first argument of the max function will be very close to one. On singular E-surfaces, the first argument is zero, such that only the second argument might give a non-zero contribution. 
If the deformation deforms in the wrong direction, the argument of $h_{\theta}$ is either negative or zero, such that $c_{ij} = 0$ as soon as the deformation deforms in the wrong direction on singular surfaces. 
Furthermore, we multiply each contribution by a falloff function $g$ defined as
\begin{align}
    \label{eq:UV_falloff_function}
    g\left(\vec{k},M_g,\gamma_g\right) \equiv \frac{\gamma_g M_g^2}{\vec{k}^2+M_g^2}.
\end{align}
The second part of eq. \eqref{eq:deformation_direction_vector} is the soft insertion part of the deformation. 
This is required since the first part of the deformation is not guaranteed to give a contribution on intersections of multiple E-surfaces. 
It is again defined with an anti-selection principle.
The basic idea is to test a number of pre-defined directions \cite{Becker:2012nk}. 
The number $N_{\mathrm{soft}}$ of pre-defined directions is a free technical parameter of the algorithm. 
As this number goes to infinity, the method is exact.
In practice it turns out that the Cartesian coordinate directions already work sufficiently well.
Thus we use $N_{\mathrm{soft}}=3$ and the standard basis,
\begin{align}
\label{eq:vector_kappa_defintion}
&\Vec{\kappa}_a = E_{\text{soft}} \Vec{e}_a, \;\; a=1,2,3\\
&\Vec{e}_1 = \begin{pmatrix}1 \\ 0 \\ 0\end{pmatrix},\;\;\Vec{e}_2 = \begin{pmatrix}0 \\ 1 \\ 0\end{pmatrix},\;\;\Vec{e}_3 = \begin{pmatrix}0 \\ 0 \\ 1\end{pmatrix},
\end{align}
with the technical parameter $E_{\text{soft}} = 0.03\sqrt{s}$.
The coefficients are defined as
\begin{align}
\label{eq:deformation_for_intersections_2}
c_{a}\left(\Vec{k},M_E,\gamma_E\right) = \prod_{\{i,l\} \in \mathcal{E}} d_{a,i,l}^+\left(\Vec{k},M_E(q_{i}^0-q_{l}^0),\gamma_E\right) - \prod_{\{i,l\} \in \mathcal{E}} d_{a,i,l}^-\left(\Vec{k},M_E(q_{i}^0-q_{l}^0),\gamma_E\right)
\end{align}
with
\begin{align}
\label{eq:deformation_for_intersections_3}
d_{a,i,l}^{\pm}\left(\Vec{k},M_E,\gamma_E\right) &= \text{max}\Bigg[h_{\delta}\left(x_{il},\sqrt{\gamma_E}M_E\right),h_{\theta}\left(\pm \Vec{\kappa}_a\cdot\Vec{b}_{il},\sqrt{\gamma_E}M_E\right)\Bigg].
\end{align}
The idea of this algorithm is the following: for each point $\Vec{k}$ and hence especially on the intersection of multiple singularities each factor $d_{a,i,l}^\pm$ checks whether $\pm\Vec{\kappa}_a$ satisfies the $i\delta$-prescription of the dual propagator $l$ when cutting the propagator $i$. 
Thus the prefactor $c_{a}$ is only non-zero on singularities if the corresponding deformation direction $\pm\Vec{\kappa}_a$ fulfils the $i\delta$-prescription of all singular $x_{ij}$ simultaneously. 

We remark that with a finite number of pre-defined directions one may construct counter-examples where no correct deformation is found.
This is unproblematic if the probability for these events can be neglected.
We further remark that one may replace the soft deformation algorithm with the computationally more expensive algorithm of ref.~\cite{Capatti:2019edf}.
The latter guarantees a correct deformation.

\subsection{Deformation magnitude}
\label{ssec:deformation_magnitude}
The scaling parameter $\lambda$ in eq. \eqref{eq:deformation_direction} should in principle be as high as possible to avoid singularities and therefore improve numerical convergence. 
However, it has three constraints that limit its magnitude \cite{Gong:2008ww,Capatti:2019edf}, namely the continuity constraint $\lambda_{\text{C}}$, the complex pole constraint $\lambda_{\text{CP}}$ and the expansion validity constraint $\lambda_{\text{exp}}$. 
We set 
\begin{align}
    \label{eq:scaling_paramter}
    \lambda = \text{min}(\lambda_{\text{max}},\lambda_{\text{C}},\lambda_{\text{CP}},\lambda_{\text{exp}}),
\end{align}
with $\lambda_{\text{max}}$ as a technical parameter of the deformation which we set to $1$. We explain the other parameter in the following.

\subsubsection{Continuity constraints}
\label{ssec:continuity_constraints}
We start by defining 
\begin{align}
	\begin{split}
	a_i &= \Vec{k}_i^2 + m^2\\
	b_i &= \Vec{\kappa}\cdot\Vec{k}_i\\
	c_i &= \Vec{\kappa}^2,
	\end{split}
\end{align}
and expand energy factors for small values of $\lambda$
\begin{align}
	\begin{split}
	E_i'&=\sqrt{a_i + 2i\lambda b_i-\lambda^2c_i}\\
	&\underset{\lambda \; \text{small}}{=} \sqrt{a_i} + \lambda \frac{ib_i}{\sqrt{a_i}} - \lambda^2 \frac{a_ic_i-b_i^2}{2a_i^{3/2}} + \lambda^3 ib_i\frac{a_ic_i-b_i^2}{2a_i^{5/2}} + \mathcal{O}(\lambda^4),
	\end{split}
	\label{eq:transformed_energy_factor_expansion}
\end{align}
for $\Vec{\kappa},m\neq0$.
We restrict the real part to be greater or equal to zero in order to avoid cutting the negative real axis and therefore violate continuity
\begin{align}
	\label{eq:continuity}
	\begin{split}
	&a_i - \lambda^2c_i \geq 0\\ 
	&\Leftrightarrow \lambda \leq \epsilon_{cc}\frac{\sqrt{a_i}}{\sqrt{c_i}} \equiv \lambda_{\text{C},i}, \;\;\; \epsilon_{cc} \in (0,1), \; \forall i \in \{1,...,N\},
	\end{split}
\end{align}
with $\epsilon_{cc}=0.95$.
Therefore
\begin{align}
    \label{eq:lambda_continuity}
    \lambda_{\text{C}} = \text{min}\left(\lambda_{\text{C},1},...,\lambda_{\text{C},N}, \lambda_{\text{C},UV} \right),
\end{align}
where $\lambda_{\text{C},UV}$ is the value for the energy appearing in the local UV subtraction term.

\subsubsection{Complex poles constraints}
\label{ssec:Complex_poles_constraints}
We calculate complex poles of E-surfaces by expanding their two energy factors according to eq. \eqref{eq:transformed_energy_factor_expansion} up to second order.
We then set real and imaginary part separately to zero as demonstrated in the following
\begin{align}
	\label{eq:complex_poles_factorized}
	\begin{split}
	&E_i + E_j - q_{ij}^0 - \lambda^2 \left( \frac{a_ic_i-b_i^2}{2a_i^{3/2}} +  \frac{a_jc_j-b_j^2}{2a_j^{3/2}}\right) + i\lambda \left( \frac{b_i}{\sqrt{a}_i} +  \frac{b_j}{\sqrt{a}_j}\right)\\
	&= A - \lambda^2C + 2i\lambda B = 0\\
	&\Rightarrow \lambda = i \frac{B}{C} \pm \sqrt{\frac{A}{C}-\left(\frac{B}{C}\right)^2},
	\end{split}
\end{align}
with the definitions
\begin{align}
    \begin{split}
        A &= E_i + E_j - q_{ij}^0 \\
        B &= \frac{1}{2}\left(\frac{b_i}{\sqrt{a}_i} +  \frac{b_j}{\sqrt{a}_j}\right)\\
        C &= \frac{a_ic_i-b_i^2}{2a_i^{3/2}} +  \frac{a_jc_j-b_j^2}{2a_j^{3/2}}.
    \end{split}
\end{align}
We avoid these complex poles by defining 
\begin{align}
\begin{split}
        \Rightarrow \lambda_{ij}^2 = 
        \begin{cases} 
        A/4C : 2\left(\frac{B}{C}\right)^2 < \frac{A}{C} \\
        \left(\frac{B}{C}\right)^2 - \frac{A}{4C} : 0 < \frac{A}{C} < 2 \left(\frac{B}{C}\right)^2 \\
        \left(\frac{B}{C}\right)^2 - \frac{A}{2C} : \frac{A}{C} < 0 
        \end{cases}
        ,
\end{split}
\end{align}
along the lines of \cite{Gong:2008ww}.
For the special case of $i=j$, which either appears in the prefactor of the causal representation or the UV subtraction term, we recall that in the previous section \ref{ssec:continuity_constraints} in eq. \eqref{eq:continuity} we constrained the real part of energy factors to be greater or equal to zero, resulting in the constraint
\begin{align}
	\label{eq:continuity_recall}
	\begin{split}
	\lambda \leq \epsilon_{cc}\frac{\sqrt{a_i}}{\sqrt{c_i}} \equiv \lambda_{\text{C},i}, \;\;\; \epsilon_{cc} \in (0,1), \; \forall i \in \{1,...,N\}.
	\end{split}
\end{align}
This already guarantees that we will not have any complex poles, since the real part is always greater than zero for $\epsilon_{cc} \in (0,1)$. 
Therefore there is no extra factor included for complex poles of energy factors.
Finally, the restrictions from complex poles are summarized to
\begin{align}
    \label{eq:complex_poles_lambda}
    \lambda_{\text{CP}} = \text{min}\left(\lambda_{\text{CP,UV}},\underset{\{i,j\}\in \mathcal{E}^{\text{all}}}{\text{min}}\left(\lambda_{ij}\right)\right),
\end{align}
where now $\mathcal{E}^{\text{all}}$ is the set of indices for all possible E-surfaces, not only the singular ones as in $\mathcal{E}$. 
This is due to the fact that even though E-surfaces might not have a real solution, they always have a complex solution.

\subsubsection{Expansion validity constraints}
\label{ssec:Expansion_validity_constraints}
In order to constrain the magnitude due to the expansion of square root factors we recall the form of the energy factor after applying contour deformation in eq. \eqref{eq:transformed_energy_factor},
\begin{align}
    E_i' = \sqrt{a_i + 2i\lambda b_i - \lambda^2 c_i} = \sqrt{a_i}\sqrt{ 1 + 2i\lambda \frac{b_i}{a_i} - \lambda^2 \frac{c_i}{a_i}}.
\end{align}
and its expansion for small values of $\lambda$ in eq. \eqref{eq:transformed_energy_factor_expansion}
\begin{align}
	\begin{split}
	\frac{1}{\sqrt{a_i}}E_i'&\underset{\lambda\; \text{small}}{=}1 + \lambda \frac{ib_i}{a_i} - \lambda^2 \frac{a_ic_i-b_i^2}{2a_i^{2}} + \lambda^3 ib_i\frac{a_ic_i-b_i^2}{2a_i^{3}} + \mathcal{O}(\lambda^4).
	\end{split}
\end{align}
Using the definitions
\begin{align}
	\frac{1}{\sqrt{a_i}}E_i' = z_0 + i \lambda z_1 - \lambda^2 z_2 + i \lambda^3 z_3 + \mathcal{O}(\lambda^4),
\end{align}
we have the relation
\begin{align}
	\frac{z_2}{z_0} = \frac{z_3}{z_1}.
\end{align}
We require 
\begin{align}
	\lambda^2\frac{z_2}{z_0} = \lambda^2\frac{a_ic_i-b_i^2}{2a_i^2} < \epsilon_{th},
\end{align}
with $\epsilon_{th} = 0.95$ in our implementation.
This guarantees that the contributions of higher orders fall off and is implemented by the constraint
\begin{align}
	\lambda^2_{\text{exp}} = \epsilon_{th} \text{min}_{j\in \{1,\dots,N\}} \left\{\frac{2a_j}{a_jc_j-b_j^2}\right\}.
\end{align}

\subsection{Parametrisation}
\label{ssec:Parametrisation}
Since Monte Carlo integrators usually provides random numbers in a hypercube, we map the numbers $(x,y,z)\in [0,1]^3$ to $(k_1,k_2,k_3)\in(-\infty,\infty)$ using the following transformation
\begin{align}
    \begin{split}
        k_1 &= \frac{x}{x-1} \cos{(2\pi z)} 2 \sqrt{y-y^2}\\
        k_2 &= \frac{x}{x-1} \sin{(2\pi z)} 2 \sqrt{y-y^2}\\
        k_3 &= \frac{x}{x-1} (1-2y)\\
        J &= 4 \pi \frac{x^2}{(x-1)^4},
    \end{split}
\end{align}
that has been used in \cite{Capatti:2019edf}. $J$ denotes the Jacobian of this transformation.

\section{Validation}
\label{sec:validation}

In order to validate our method we compare the Monte Carlo integrated result for one-loop amplitudes with up to seven legs obtained from the causal loop-tree duality representation with analytical results.
We used the RAMBO algorithm \cite{Kleiss:1985gy} to generate configurations of external momenta that conserve overall momentum 
and satisfy the on-shell conditions.
By convention, we take the momenta of particles $1$ and $2$ to have a negative energy component while the momenta of all other particles have a positive energy component.
The centre of mass energy of the two incoming particles is randomly generated.
We set the unit of energy equal to one (\textit{i.e.} $1 \; \mathrm{GeV}=1$).
We consider renormalised amplitudes in the $\overline{\text{MS}}$-scheme and use $\mu_{\text{UV}}^2 = 1$.
\newline
The reference values were generated using FeynArts \cite{Hahn:2000kx} to output all possible diagrams of the amplitude and using the Loop Tools package \cite{Hahn:1998yk} for $4,5$-point functions and the Collier package \cite{Denner:2016kdg} for $4-7$-point functions to evaluate the resulting integrals. 
For integration we use the Monte Carlo integrator Vegas provided by the CUBA library \cite{Hahn:2005pf,Hahn:2014fua}. 
We integrated $500$ $4$-point amplitudes, $50$ $5$-point amplitudes, $4$ $6$-point amplitudes and $4$ $7$-point amplitudes.
The number of evaluations was set to $50\cdot10^{6}$ Monte Carlo points.
We observed that the Monte Carlo error reported by the Monte Carlo integrator is
orders of magnitudes higher than the relative error compared to the reference value in some cases. 
We therefore conclude that the Monte Carlo error estimate is too high and do not report it. 
We first give explicit results for one specific configuration of external momenta for each amplitude in table \ref{tab:specific_integration_results}. 
For efficiency reasons we ignored factors of $i$ stemming from propagators and vertices which would result in an overall factor of $(-1)^N$ in the amplitude.  
\begin{table}[ht!]
    \centering
    \begin{tabular}{c|c|c|c}
        Amplitude & numerical value & reference value & relative error\\\hline\hline
        4-point real & $3.7715\cdot10^{-11}$ & $ 3.7728\cdot10^{-11}$ & $3.2 \cdot10^{-4}$\\\hline
        4-point imag. & $5.4217\cdot10^{-11}$ & $5.4200\cdot10^{-11}$ & $3.3 \cdot10^{-4}$ \\\hline
        5-point real & $1.4172\cdot10^{-15}$ & $1.4153\cdot10^{-15}$ &  $1.4 \cdot10^{-3}$\\\hline
        5-point imag. & $-1.2211\cdot10^{-14}$ & $-1.2209\cdot10^{-14}$ &  $2.1 \cdot10^{-4}$\\\hline
        6-point real & $1.378\cdot10^{-13}$ & $1.398\cdot10^{-13}$ & $1.4 \cdot10^{-2}$\\\hline
        6-point imag. & $9.243\cdot10^{-13}$ & $9.209\cdot10^{-13}$ & $3.7 \cdot10^{-3}$\\\hline
        7-point real & $-3.088\cdot10^{-21}$ & $-3.051\cdot10^{-21}$ &  $1.2 \cdot10^{-2}$\\\hline
        7-point imag. & $-6.613\cdot10^{-21}$ & $-6.655\cdot10^{-21}$ & $6.4 \cdot10^{-3}$
    \end{tabular}
    \caption{\small Integration results separated into real and imaginary (imag.) part}
    \label{tab:specific_integration_results}
\end{table}
The external momenta to reproduce these results are given in appendix \ref{sec:external_momenta}.
Then we summarised the remaining results in table \ref{tab:integration_results}, by giving the average relative error as well as its standard deviation.
\begin{table}[ht!]
    \centering
    \begin{tabular}{c|c|c}
        Amplitude & average relative error & standard deviation\\\hline\hline
        4-point real & $1.5\cdot10^{-3}$ & $8.8\cdot10^{-3}$ \\\hline
        4-point imag. & $5.6\cdot10^{-4}$ & $1.0\cdot10^{-3}$ \\\hline
        5-point real & $3.2\cdot10^{-3}$ & $8.7\cdot10^{-3}$ \\\hline
        5-point imag. & $2.4\cdot10^{-3}$ & $9.7\cdot10^{-3}$ \\\hline
        6-point real & $4.9\cdot10^{-3}$ & $6.4\cdot10^{-3}$ \\\hline
        6-point imag. & $1.5\cdot10^{-3}$ & $1.5\cdot10^{-3}$ \\\hline
        7-point real & $1.5\cdot10^{-2}$ & $1.2\cdot10^{-2}$ \\\hline
        7-point imag. & $7.9\cdot10^{-3}$ & $6.6\cdot10^{-3}$
    \end{tabular}
    \caption{\small Average error of integration results separated into real and imaginary part}
    \label{tab:integration_results}
\end{table}
The average relative error is obtained as the relative error to the reference values.
We want to stress that the code itself as well as the contour deformation have not been optimised yet using methods mentioned in \cite{Becker:2012aqa,Capatti:2019edf}.
For 1 million Monte Carlo integration points the $4$-point amplitudes usually take a few seconds each, $5,6$-point a few minutes and $7$-point a few hours on a single standard PC. 
A general remark of table \ref{tab:integration_results} is that the average error of the imaginary part is usually smaller than the one of the real part. 
Furthermore more external particles result in a slower convergence. 

\section{Conclusions}
\label{sec:conclusions}

In this paper we considered renormalised one-loop amplitudes in massive $\phi^3$-theory
within causal loop-tree duality.
We derived a recurrence relation for the integrand of the one-loop amplitude, which only contains
causal terms.
The recurrence relation is directly based on the causal representation and avoids Feynman diagrams.
The resulting integrand can be integrated numerically by Monte Carlo methods.
We have verified up to seven points that our results agree with analytic results.

Our results are an important step towards the application of loop-tree duality towards
the calculations of realistic observables and cross-sections within this framework.

We expect the generalisation to other theories (e.g. Yang-Mills, QCD) to be unproblematic,
as these theories differ only by numerator terms.
The diagrams of $\phi^3$-theory are the most general ones: Any other theory will only add
diagrams which are obtained by pinching some propagators.
These additional diagrams do not introduce new causal terms.

\subsection*{Acknowledgements}

This work has been partially supported
by the Cluster of Excellence Precision Physics, Fundamental Interactions, and Structure of
Matter (PRISMA EXC 2118/1) funded by the German Research Foundation (DFG) within
the German Excellence Strategy (Project ID 39083149).

\appendix
\section{Derivation of the number of causal terms per diagram}\label{sec:app_num_causal_terms}
In this appendix we derive the number of causal terms for a single diagram given in equation \eqref{eq:number_of_causal_terms_diagram}. We start by calculating the number of terms of $\mathcal{F}$, \textit{c.f.} equation \eqref{eq:pprod} by induction in $N_L$, the number of elements of the set $A_L$. At this point we remind the reader of the relation $N = N_L + N_R$ since it will be important in the following.\\
The case $N_L = 1$ is special since there is no sum, as can be seen in the first line of equation \eqref{eq:pprod_first_term}. The hypothesis that we want to prove is that the number of terms, $N_{\text{terms}}^\mathcal{F}(N_L, N_R)$ depending on the cardinalities $N_L$ and $N_R$, is given by
\begin{align}
    \label{eq:hypothesis_num_terms_in_F}
    N_{\text{terms}}^\mathcal{F}(N_L, N_R) \coloneqq \sum_{m_1=1}^{N_R} \sum_{m_2=1}^{m_1} \dots \sum_{m_{N_L-1}=1}^{m_{N_L-2}} 1 = \binom{N-2}{N_L-1},
\end{align}
for $N_L \geq 1$.\\
For $N_L = 1$ we trivially obtain $N_{\text{terms}}^\mathcal{F}(1, N_R) = 1 = \binom{1+N_R-2}{0}$.\\
Now suppose equation \eqref{eq:hypothesis_num_terms_in_F} holds for $N_L$. To show that it still holds for $(N_L+1)$, let us replace the last $(N_L-1)$ sums using equation \eqref{eq:hypothesis_num_terms_in_F},
\begin{align*}
    \sum_{m_1=1}^{N_R} \sum_{m_2=1}^{m_1} \dots \sum_{m_{N_L}=1}^{m_{N_L-1}} 1 = \sum_{m=1}^{N_R} \binom{m+N_L-2}{N_L-1}.
\end{align*}
Now we can shift the index of the sum on the right-hand side and use the identity
\begin{align*}
    \sum_{j=k}^n \binom{j}{k} = \binom{n+1}{k+1}
\end{align*}
to obtain
\begin{align*}
   \sum_{m=1}^{N_R} \binom{m+N_L-2}{N_L-1} = \sum_{m=N_L-1}^{N-2} \binom{m}{N_L-1} = \binom{N-1}{N_L},
\end{align*}
This proves equation \eqref{eq:hypothesis_num_terms_in_F}.\\
We are now able to compute the total number of terms generated by the expression on the right-hand-side of equation \eqref{eq:pprod},
\begin{align}
    \label{eq:num_tot_terms}
    N_{\text{terms}}^{\text{tot}} (N) \coloneqq \sum_{\substack{N_L=1,\\N_R=N-N_L}}^{N-1} \sum_{A_L \sqcup A_R} \binom{N-2}{N_L-1}.
\end{align}
Since the expression on the right-hand-side only depends on the cardinalities of the sets $A_L$ and $A_R$ but not on the particular elements in them, we can simply count the number of all partitions of the set $\{1, \dots N\}$ into two disjoint subsets, $\sum_{A_L \sqcup A_R}$, which is given by
\begin{align*}
    \sum_{A_L \sqcup A_R} 1 = \binom{N}{N_L}.
\end{align*}
Plugging this into equation \eqref{eq:num_tot_terms}, we arrive at
\begin{align*}
    N_{\text{terms}}^{\text{tot}}(N) &\coloneqq \sum_{\substack{N_L=1,\\N_R=N-N_L}}^{N-1} \binom{N}{N_L} \binom{N-2}{N_L-1}\\
    &= \sum_{\substack{N_L=1,\\N_R=N-N_L}}^{N-1} \binom{N}{N_L} \binom{2(N-1) - N}{N-1-N_L}\\
    &= \binom{2(N-1)}{N-1},
\end{align*}
where we used the Chu-Vandermonde identity to get to the last line.

\section{External momenta}\label{sec:external_momenta}
In this appendix we give the external momenta used for the $N$-point amplitudes calculated in table \ref{tab:specific_integration_results}. 
All momenta are on-shell $p_i^2=m^2$ $\forall$ $i =1,\dots, N$ and the $N$'th momentum is given by momentum conservation. 

\subsection*{$4$-point amplitude}
\begin{align*}
\begin{split}
    p_1 = &(-141.10983711348885,	0.0,	0.0,	133.14844423055888) \\
    p_2 = &(-141.10983711348885,	0.0,	0.0,	-133.14844423055888) \\
    p_3 = &(141.10983711348888,	-49.09398540018926,	\\
    &\phantom{(} -123.76633930886697,	-0.4266762551018309	) \\
    m = &46.72769980618679
\end{split}
\end{align*}
\subsection*{$5$-point amplitude}
\begin{align*}
\begin{split}
    p_1 = &(-219.15224144836964,	0.0,	0.0,	213.5527212086298) \\
    p_2 = &(-219.15224144836964,	0.0,	0.0,	-213.5527212086298) \\
    p_3 = &(159.3685658563847,	100.88986032846607,		\\
    &\phantom{(} 103.57375828958654,	-45.48749568353928) \\
    p_4 = &(139.654501422648,	-107.72894459349114,		\\
    &\phantom{(} -35.08814424880286,	-65.1439651949617) \\
    m = &49.223370427407076
\end{split}
\end{align*}
\subsection*{$6$-point amplitude}
\begin{align*}
\begin{split}
    p_1 = &(-71.35064013734655,	0.0,	0.0,	70.92438940359767) \\
    p_2 = &(-71.35064013734655,	0.0,	0.0,	-70.92438940359767) \\
    p_3 = &(29.192886202212556,	-4.118173727335252,			\\
    &\phantom{(} 24.047177858045522,	-14.012624711108641) \\
    p_4 = &(42.54868938111772,	-12.640455845618828,			\\
    &\phantom{(} 1.733152867873708,	39.836681456496976) \\
    p_5 = &(53.78230976095452,	21.065081112026018,			\\
    &\phantom{(} -27.466568023795073,	-40.41957454285356	) \\
    m = &7.787479421222914
\end{split}
\end{align*}
\subsection*{$7$-point amplitude}
\begin{align*}
\begin{split}
    p_1 = &(-332.36057016638324,	0.0,	0.0,	332.2385272592629) \\
    p_2 = &(-332.36057016638324,	0.0,	0.0,	-332.2385272592629) \\
    p_3 = &(139.74515554691573,	-39.48281923163617,			\\
    &\phantom{(} -130.247185116811,	30.40356307378165) \\
    p_4 = &(152.89374330062682,	4.2838059333724425,			\\
    &\phantom{(}	-148.2848545094504,	-35.89760527381943) \\
    p_5 = &(285.42775357977837,	-5.4421142702661465,	\\
    &\phantom{(} 280.0622653218135,	-54.06851063939918	) \\
    p_6 = &(12.720773095787127,	0.512475927638319,		\\
    &\phantom{(} -8.853020930392365,	-1.4386976974916026) \\
    m = &9.006087159214037
\end{split}
\end{align*}

\clearpage

\addcontentsline{toc}{section}{Bibliography}
\bibliographystyle{bibliography/JHEP-2}
\bibliography{bibliography/bibliography.bib}

\providecommand{\href}[2]{#2}\begingroup\raggedright\begin{thebibliography}{10}

\bibitem{Catani:2008xa}
S.~Catani, T.~Gleisberg, F.~Krauss, G.~Rodrigo and J.-C. Winter, {\normalfont
  \itshape {From loops to trees by-passing Feynman's theorem}},  {\em JHEP}
  {\normalfont \bfseries 09} (2008) 065
  [\href{http://arXiv.org/abs/0804.3170}{{\tt 0804.3170}}].

\bibitem{Bierenbaum:2010cy}
I.~Bierenbaum, S.~Catani, P.~Draggiotis and G.~Rodrigo, {\normalfont \itshape
  {A Tree-Loop Duality Relation at Two Loops and Beyond}},  {\em JHEP}
  {\normalfont \bfseries 10} (2010) 073
  [\href{http://arXiv.org/abs/1007.0194}{{\tt 1007.0194}}].

\bibitem{CaronHuot:2010zt}
S.~Caron-Huot, {\normalfont \itshape {Loops and trees}},  {\em JHEP}
  {\normalfont \bfseries 05} (2011) 080
  [\href{http://arXiv.org/abs/1007.3224}{{\tt 1007.3224}}].
%%CITATION = ARXIV:1007.3224;%%

\bibitem{Bierenbaum:2012th}
I.~Bierenbaum, S.~Buchta, P.~Draggiotis, I.~Malamos and G.~Rodrigo,
  {\normalfont \itshape {Tree-Loop Duality Relation beyond simple poles}},
  {\em JHEP} {\normalfont \bfseries 03} (2013) 025
  [\href{http://arXiv.org/abs/1211.5048}{{\tt 1211.5048}}].

\bibitem{Buchta:2014dfa}
S.~Buchta, G.~Chachamis, P.~Draggiotis, I.~Malamos and G.~Rodrigo, {\normalfont
  \itshape {On the singular behaviour of scattering amplitudes in quantum field
  theory}},  {\em JHEP} {\normalfont \bfseries 11} (2014) 014
  [\href{http://arXiv.org/abs/1405.7850}{{\tt 1405.7850}}].

\bibitem{Hernandez-Pinto:2015ysa}
R.~J. Hernandez-Pinto, G.~F.~R. Sborlini and G.~Rodrigo, {\normalfont \itshape
  {Towards gauge theories in four dimensions}},  {\em JHEP} {\normalfont
  \bfseries 02} (2016) 044 [\href{http://arXiv.org/abs/1506.04617}{{\tt
  1506.04617}}].

\bibitem{Buchta:2015wna}
S.~Buchta, G.~Chachamis, P.~Draggiotis and G.~Rodrigo, {\normalfont \itshape
  {Numerical implementation of the loop\textendash{}tree duality method}},
  {\em Eur. Phys. J. C} {\normalfont \bfseries 77} (2017), no.~5 274
  [\href{http://arXiv.org/abs/1510.00187}{{\tt 1510.00187}}].

\bibitem{Sborlini:2016gbr}
G.~F.~R. Sborlini, F.~Driencourt-Mangin, R.~Hernandez-Pinto and G.~Rodrigo,
  {\normalfont \itshape {Four-dimensional unsubtraction from the loop-tree
  duality}},  {\em JHEP} {\normalfont \bfseries 08} (2016) 160
  [\href{http://arXiv.org/abs/1604.06699}{{\tt 1604.06699}}].

\bibitem{Driencourt-Mangin:2017gop}
F.~Driencourt-Mangin, G.~Rodrigo and G.~F.~R. Sborlini, {\normalfont \itshape
  {Universal dual amplitudes and asymptotic expansions for $gg\rightarrow H$
  and $H\rightarrow \gamma \gamma $ in four dimensions}},  {\em Eur. Phys. J.
  C} {\normalfont \bfseries 78} (2018), no.~3 231
  [\href{http://arXiv.org/abs/1702.07581}{{\tt 1702.07581}}].

\bibitem{Driencourt-Mangin:2019aix}
F.~Driencourt-Mangin, G.~Rodrigo, G.~F.~R. Sborlini and W.~J. Torres~Bobadilla,
  {\normalfont \itshape {Universal four-dimensional representation of $H \to
  \gamma \gamma$ at two loops through the Loop-Tree Duality}},  {\em JHEP}
  {\normalfont \bfseries 02} (2019) 143
  [\href{http://arXiv.org/abs/1901.09853}{{\tt 1901.09853}}].
%%CITATION = ARXIV:1901.09853;%%

\bibitem{Runkel:2019yrs}
R.~Runkel, Z.~Sz\H{o}r, J.~P. Vesga and S.~Weinzierl, {\normalfont \itshape
  {Causality and loop-tree duality at higher loops}},  {\em Phys. Rev. Lett.}
  {\normalfont \bfseries 122} (2019), no.~11 111603
  [\href{http://arXiv.org/abs/1902.02135}{{\tt 1902.02135}}]. [Erratum:
  Phys.Rev.Lett. 123, 059902 (2019)].

\bibitem{Runkel:2019zbm}
R.~Runkel, Z.~Sz\H{o}r, J.~P. Vesga and S.~Weinzierl, {\normalfont \itshape
  {Integrands of loop amplitudes within loop-tree duality}},  {\em Phys. Rev.
  D} {\normalfont \bfseries 101} (2020), no.~11 116014
  [\href{http://arXiv.org/abs/1906.02218}{{\tt 1906.02218}}].

\bibitem{Baumeister:2019rmh}
R.~Baumeister, D.~Mediger, J.~Pe\v{c}ovnik and S.~Weinzierl, {\normalfont
  \itshape {Vanishing of certain cuts or residues of loop integrals with higher
  powers of the propagators}},  {\em Phys. Rev. D} {\normalfont \bfseries 99}
  (2019), no.~9 096023 [\href{http://arXiv.org/abs/1903.02286}{{\tt
  1903.02286}}].

\bibitem{Aguilera-Verdugo:2019kbz}
J.~J. Aguilera-Verdugo, F.~Driencourt-Mangin, J.~Plenter,
  S.~Ram\'\i{}rez-Uribe, G.~Rodrigo, G.~F.~R. Sborlini, W.~J. Torres~Bobadilla
  and S.~Tracz, {\normalfont \itshape {Causality, unitarity thresholds,
  anomalous thresholds and infrared singularities from the loop-tree duality at
  higher orders}},  {\em JHEP} {\normalfont \bfseries 12} (2019) 163
  [\href{http://arXiv.org/abs/1904.08389}{{\tt 1904.08389}}].

\bibitem{Driencourt-Mangin:2019yhu}
F.~Driencourt-Mangin, G.~Rodrigo, G.~F.~R. Sborlini and W.~J. Torres~Bobadilla,
  {\normalfont \itshape {Interplay between the loop-tree duality and helicity
  amplitudes}},  {\em Phys. Rev. D} {\normalfont \bfseries 105} (2022), no.~1
  016012 [\href{http://arXiv.org/abs/1911.11125}{{\tt 1911.11125}}].

\bibitem{Aguilera-Verdugo:2020set}
J.~J. Aguilera-Verdugo, F.~Driencourt-Mangin, R.~J. Hern\'andez-Pinto,
  J.~Plenter, S.~Ramirez-Uribe, A.~E. Renteria~Olivo, G.~Rodrigo, G.~F.~R.
  Sborlini, W.~J. Torres~Bobadilla and S.~Tracz, {\normalfont \itshape {Open
  Loop Amplitudes and Causality to All Orders and Powers from the Loop-Tree
  Duality}},  {\em Phys. Rev. Lett.} {\normalfont \bfseries 124} (2020), no.~21
  211602 [\href{http://arXiv.org/abs/2001.03564}{{\tt 2001.03564}}].

\bibitem{Plenter:2020lop}
J.~Plenter and G.~Rodrigo, {\normalfont \itshape {Asymptotic expansions through
  the loop-tree duality}},  {\em Eur. Phys. J. C} {\normalfont \bfseries 81}
  (2021), no.~4 320 [\href{http://arXiv.org/abs/2005.02119}{{\tt 2005.02119}}].

\bibitem{Aguilera-Verdugo:2020kzc}
J.~J. Aguilera-Verdugo, R.~J. Hernandez-Pinto, G.~Rodrigo, G.~F.~R. Sborlini
  and W.~J. Torres~Bobadilla, {\normalfont \itshape {Causal representation of
  multi-loop Feynman integrands within the loop-tree duality}},  {\em JHEP}
  {\normalfont \bfseries 01} (2021) 069
  [\href{http://arXiv.org/abs/2006.11217}{{\tt 2006.11217}}].

\bibitem{Ramirez-Uribe:2020hes}
S.~Ram\'\i{}rez-Uribe, R.~J. Hern\'andez-Pinto, G.~Rodrigo, G.~F.~R. Sborlini
  and W.~J. Torres~Bobadilla, {\normalfont \itshape {Universal opening of
  four-loop scattering amplitudes to trees}},  {\em JHEP} {\normalfont
  \bfseries 04} (2021) 129 [\href{http://arXiv.org/abs/2006.13818}{{\tt
  2006.13818}}].

\bibitem{JesusAguilera-Verdugo:2020fsn}
J.~Jes\'us Aguilera-Verdugo, R.~J. Hern\'andez-Pinto, G.~Rodrigo, G.~F.~R.
  Sborlini and W.~J. Torres~Bobadilla, {\normalfont \itshape {Mathematical
  properties of nested residues and their application to multi-loop scattering
  amplitudes}},  {\em JHEP} {\normalfont \bfseries 02} (2021) 112
  [\href{http://arXiv.org/abs/2010.12971}{{\tt 2010.12971}}].

\bibitem{TorresBobadilla:2021ivx}
W.~J. Torres~Bobadilla, {\normalfont \itshape {Loop-tree duality from vertices
  and edges}},  {\em JHEP} {\normalfont \bfseries 04} (2021) 183
  [\href{http://arXiv.org/abs/2102.05048}{{\tt 2102.05048}}].

\bibitem{Sborlini:2021owe}
G.~F.~R. Sborlini, {\normalfont \itshape {Geometrical approach to causality in
  multiloop amplitudes}},  {\em Phys. Rev. D} {\normalfont \bfseries 104}
  (2021), no.~3 036014 [\href{http://arXiv.org/abs/2102.05062}{{\tt
  2102.05062}}].

\bibitem{Bobadilla:2021pvr}
W.~J.~T. Bobadilla, {\normalfont \itshape {Lotty \textendash{} The loop-tree
  duality automation}},  {\em Eur. Phys. J. C} {\normalfont \bfseries 81}
  (2021), no.~6 514 [\href{http://arXiv.org/abs/2103.09237}{{\tt 2103.09237}}].

\bibitem{deJesusAguilera-Verdugo:2021mvg}
J.~de~Jes\'us Aguilera-Verdugo {\em et.~al.}, {\normalfont \itshape {A Stroll
  through the Loop-Tree Duality}},  {\em Symmetry} {\normalfont \bfseries 13}
  (2021), no.~6 1029 [\href{http://arXiv.org/abs/2104.14621}{{\tt
  2104.14621}}].

\bibitem{Benincasa:2021qcb}
P.~Benincasa and W.~J.~T. Bobadilla, {\normalfont \itshape {Physical
  representations for scattering amplitudes and the wavefunction of the
  universe}},  {\em SciPost Phys.} {\normalfont \bfseries 12} (2022), no.~6 192
  [\href{http://arXiv.org/abs/2112.09028}{{\tt 2112.09028}}].

\bibitem{Capatti:2019ypt}
Z.~Capatti, V.~Hirschi, D.~Kermanschah and B.~Ruijl, {\normalfont \itshape
  {Loop-Tree Duality for Multiloop Numerical Integration}},  {\em Phys. Rev.
  Lett.} {\normalfont \bfseries 123} (2019), no.~15 151602
  [\href{http://arXiv.org/abs/1906.06138}{{\tt 1906.06138}}].

\bibitem{Capatti:2019edf}
Z.~Capatti, V.~Hirschi, D.~Kermanschah, A.~Pelloni and B.~Ruijl, {\normalfont
  \itshape {Numerical Loop-Tree Duality: contour deformation and subtraction}},
   {\em JHEP} {\normalfont \bfseries 04} (2020) 096
  [\href{http://arXiv.org/abs/1912.09291}{{\tt 1912.09291}}].

\bibitem{Capatti:2020ytd}
Z.~Capatti, V.~Hirschi, D.~Kermanschah, A.~Pelloni and B.~Ruijl, {\normalfont
  \itshape {Manifestly Causal Loop-Tree Duality}},
  \href{http://arXiv.org/abs/2009.05509}{{\tt 2009.05509}}.

\bibitem{Capatti:2020xjc}
Z.~Capatti, V.~Hirschi, A.~Pelloni and B.~Ruijl, {\normalfont \itshape {Local
  Unitarity: a representation of differential cross-sections that is locally
  free of infrared singularities at any order}},  {\em JHEP} {\normalfont
  \bfseries 04} (2021) 104 [\href{http://arXiv.org/abs/2010.01068}{{\tt
  2010.01068}}].

\bibitem{Kermanschah:2021wbk}
D.~Kermanschah, {\normalfont \itshape {Numerical integration of loop integrals
  through local cancellation of threshold singularities}},  {\em JHEP}
  {\normalfont \bfseries 01} (2022) 151
  [\href{http://arXiv.org/abs/2110.06869}{{\tt 2110.06869}}].

\bibitem{Capatti:2022tit}
Z.~Capatti, V.~Hirschi and B.~Ruijl, {\normalfont \itshape {Local Unitarity:
  cutting raised propagators and localising renormalisation}},
  \href{http://arXiv.org/abs/2203.11038}{{\tt 2203.11038}}.

\bibitem{Buchta:2015jea}
S.~Buchta, G.~Chachamis, P.~Draggiotis, I.~Malamos and G.~Rodrigo, {\normalfont
  \itshape {Towards a Numerical Implementation of the Loop-Tree Duality
  Method}},  {\em Nucl. Part. Phys. Proc.} {\normalfont \bfseries 258-259}
  (2015) 33--36 [\href{http://arXiv.org/abs/1509.07386}{{\tt 1509.07386}}].

\bibitem{Berends:1987me}
F.~A. Berends and W.~T. Giele, {\normalfont \itshape {Recursive Calculations
  for Processes with n Gluons}},  {\em Nucl. Phys. B} {\normalfont \bfseries
  306} (1988) 759--808.

\bibitem{Becker:2010ng}
S.~Becker, C.~Reuschle and S.~Weinzierl, {\normalfont \itshape {Numerical NLO
  QCD calculations}},  {\em JHEP} {\normalfont \bfseries 12} (2010) 013
  [\href{http://arXiv.org/abs/1010.4187}{{\tt 1010.4187}}].

\bibitem{Gong:2008ww}
W.~Gong, Z.~Nagy and D.~E. Soper, {\normalfont \itshape {Direct numerical
  integration of one-loop Feynman diagrams for N-photon amplitudes}},  {\em
  Phys. Rev. D} {\normalfont \bfseries 79} (2009) 033005
  [\href{http://arXiv.org/abs/0812.3686}{{\tt 0812.3686}}].

\bibitem{Becker:2012aqa}
S.~Becker, C.~Reuschle and S.~Weinzierl, {\normalfont \itshape {Efficiency
  Improvements for the Numerical Computation of NLO Corrections}},  {\em JHEP}
  {\normalfont \bfseries 07} (2012) 090
  [\href{http://arXiv.org/abs/1205.2096}{{\tt 1205.2096}}].

\bibitem{Becker:2012nk}
S.~Becker and S.~Weinzierl, {\normalfont \itshape {Direct contour deformation
  with arbitrary masses in the loop}},  {\em Phys. Rev. D} {\normalfont
  \bfseries 86} (2012) 074009 [\href{http://arXiv.org/abs/1208.4088}{{\tt
  1208.4088}}].

\bibitem{Becker:2012bi}
S.~Becker and S.~Weinzierl, {\normalfont \itshape {Direct numerical integration
  for multi-loop integrals}},  {\em Eur. Phys. J. C} {\normalfont \bfseries 73}
  (2013), no.~2 2321 [\href{http://arXiv.org/abs/1211.0509}{{\tt 1211.0509}}].

\bibitem{Kleiss:1985gy}
R.~Kleiss, W.~J. Stirling and S.~D. Ellis, {\normalfont \itshape {A New Monte
  Carlo Treatment of Multiparticle Phase Space at High-energies}},  {\em
  Comput. Phys. Commun.} {\normalfont \bfseries 40} (1986) 359.

\bibitem{Hahn:2000kx}
T.~Hahn, {\normalfont \itshape {Generating Feynman diagrams and amplitudes with
  FeynArts 3}},  {\em Comput. Phys. Commun.} {\normalfont \bfseries 140} (2001)
  418--431 [\href{http://arXiv.org/abs/hep-ph/0012260}{{\tt hep-ph/0012260}}].

\bibitem{Hahn:1998yk}
T.~Hahn and M.~Perez-Victoria, {\normalfont \itshape {Automatized one loop
  calculations in four-dimensions and D-dimensions}},  {\em Comput. Phys.
  Commun.} {\normalfont \bfseries 118} (1999) 153--165
  [\href{http://arXiv.org/abs/hep-ph/9807565}{{\tt hep-ph/9807565}}].

\bibitem{Denner:2016kdg}
A.~Denner, S.~Dittmaier and L.~Hofer, {\normalfont \itshape {Collier: a
  fortran-based Complex One-Loop Library in Extended Regularizations}},  {\em
  Comput. Phys. Commun.} {\normalfont \bfseries 212} (2017) 220--238
  [\href{http://arXiv.org/abs/1604.06792}{{\tt 1604.06792}}].

\bibitem{Hahn:2005pf}
T.~Hahn, {\normalfont \itshape {The CUBA library}},  {\em Nucl. Instrum. Meth.
  A} {\normalfont \bfseries 559} (2006) 273--277
  [\href{http://arXiv.org/abs/hep-ph/0509016}{{\tt hep-ph/0509016}}].

\bibitem{Hahn:2014fua}
T.~Hahn, {\normalfont \itshape {Concurrent Cuba}},  {\em J. Phys. Conf. Ser.}
  {\normalfont \bfseries 608} (2015), no.~1 012066
  [\href{http://arXiv.org/abs/1408.6373}{{\tt 1408.6373}}].

\end{thebibliography}\endgroup
\end{document}